\documentclass[aip,jcp,reprint]{revtex4-1}
\usepackage{graphicx} 
\usepackage{subcaption}
\usepackage{amsmath} 
\usepackage{amssymb}
\usepackage{color}
\usepackage{textcomp}
\begin{document}

% \preprint{}

\title{Quantum reactive scattering of O($^3$P)+H$_2$ at collision energies up to 4.4 eV} %Title of paper



\author{Marko Gacesa}
\email[]{gacesa@phys.uconn.edu}
%\thanks{}
\affiliation{Department of Physics, University of Connecticut, Storrs, CT}

\author{Vasili Kharchenko}
\email[]{kharchenko@phys.uconn.edu}
\affiliation{Department of Physics, University of Connecticut, Storrs, CT}
\affiliation{ITAMP, Harvard-Smithsonian Center for Astrophysics, Cambridge, MA}


\date{\today}

\begin{abstract}
We report the results of quantum scattering calculations for the O($^3$P)+H$_2$ reaction for a range of collision energies from 0.4 to 4.4 eV, important for astrophysical and atmospheric processes. The total and state-to-state reactive cross sections are calculated using a fully quantum time-independent coupled-channel approach on recent potential energy surfaces of $^{3}A'$ and $^{3}A''$ symmetry. A larger basis set than in the previous studies was used to ensure convergence at higher energies. Our results agree well with the published data at lower energies and indicate the breakdown of reduced dimensionality approach at collision energies higher than 1.5 eV. Differential cross sections and momentum transfer cross sections are also reported. 
\end{abstract}

\pacs{}% insert suggested PACS numbers in braces on next line

\maketitle %\maketitle must follow title, authors, abstract and \pacs

\section{Introduction}
The O($^{3}P$)+H$_2$ reaction represents a prototype chemical process of considerable interest in combustion\cite{irvin2008combustion}, as well as in atmospheric\cite{2001GeoRL..28.2157R,2004GeoRL..31.4106B,2012GeoRL..3910203G} and interstellar\cite{1987ApJ...317..432G} chemistry. 
Owing to its relative simplicity, as well as to the fact that the reaction has an energy barrier of about 0.56 eV below which it proceeds via an abstraction mechanism\cite{Rogers_Kupperman_PES_2000}, it is suitable for benchmarking theoretical methods and experiments. Extensive literature about the topic has been published. 
Reaction rates and related quantities obtained using different experimental techniques have been reported \cite{1967JChPh..46..490W,1969JChPh..50.2512W,1.436078,doi:10.1139/v75-508,1985JChPh..82.1291P,Natarajan1987267,1453395,1989CPL...161..219S,doi:10.1021/j100338a029,Davidson1990445,Yang199369,Ryu1995279,10.1063/1.475858,KIN:KIN4,Han2000243,2003JChPh.118.1585G,2013NatCh...5..315L}. On the theory side, several sets of chemically accurate global potential energy surfaces (PESs) have been constructed\cite{1977JChPh..66.4116J,1981JChPh..74.4984S,1.452544,1.454396,doi:10.1021/jp970676p,2004JChPh.121.8861B_PES,Rogers_Kupperman_PES_2000,Zhai201225} and different aspects of the process have been investigated\cite{1.452544,1.454396,1992JPhB...25..285J,doi:10.1021/jp970676p,Rogers_Kupperman_PES_2000,2004JChPh.121.8861B_PES,Zhai201225,2001GeoRL..28.2157R,1981JChPh..74.4984S,10.1063/1.443394,doi:10.1021/j100409a017,QUA:QUA560290538,10.1063/1.453205,10.1063/1.473174,2003JChPh.119..195B,2003JChPh.119..195B,2003JChPh.11912360M,2004JChPh.121.6346B,2004JChPh.120.4316B,2006JChPh.124g4308W,2006CPL...418..250W,B608871F,2010ChJCP..23..149W,2010CP....368...58W,2011ChPhB..20g8201L,10.1063/1.4795497,JCC:JCC21940}.
Particularly relevant to the present study are recent time-dependent and time-independent quantum mechanical (QM) \cite{2003JChPh.119..195B,2004JChPh.121.6346B} and quasi-classical trajectories (QCT) calculations \cite{2010ChJCP..23..149W,2010CP....368...58W,2011ChPhB..20g8201L,10.1063/1.4795497,JCC:JCC21940} that have demonstrated good agreement with crossed molecular beam experiments\cite{2003JChPh.118.1585G,2012Sci...336.1687J,2013NatCh...5..315L} for temperatures ranging from thermal to about 3500 K. 

Majority of the existing studies were done for the temperatures corresponding to the collision energies of the reactants smaller than 1 eV in the center-of-mass frame.
Atmospheric and interstellar chemistry, however, often involve processes leading to collisions of hot oxygen atoms with H$_2$ molecules at energies between 1 and 5 eV\cite{2008SSRv..139..355J}. While semi-classical approaches and mass-scaling of known cross sections have been used to obtain approximate cross sections for such processes, quantum mechanical methods, if computationally feasible, offer many advantages, such as state-to-state resolution and higher accuracy\cite{2011GeoRL..3802203B,2012GeoRL..3910203G}. For example, QM scattering calculation of suprathermal collisions of atmospheric gases and hot O atoms in planetary atmospheres has been used to estimate the reactions'  contribution to formation of extended planetary coronae\cite{2007JGRE..11209009C}, and predict a new escape mechanism of molecular hydrogen from the martian atmosphere\cite{2012GeoRL..3910203G}.

In this article, we report the results of state-to-state quantum scattering calculations of O($^{3}P$)+H$_2$ reactive collision in the energy interval from 0.4 to 4.4 eV, using recent chemically accurate $^3A'$ and $^3A''$ PESs \cite{Rogers_Kupperman_PES_2000,2004JChPh.121.8861B_PES}. We compare our results with existing QM calculations and QCT calculations, as well as with the results of molecular beam measurements\cite{2003JChPh.118.1585G}. We also report momentum transfer (diffusion) cross sections and differential cross sections for the reaction of oxygen atoms with the hydrogen molecule in one of the two lowest vibrational levels.

This paper is organized as follows. In Sec. II we describe the methods used in the calculation and discuss their convergence. In Sec. III we present total and selected state-to-state reactive cross sections for neutral and vibrationally excited hydrogen molecule, as well as selected differential cross sections. The summary and conclusions are given in Sec. IV.

\section{Methods}
We carried out quantum reactive scattering calculations using computer program ABC\cite{2000CoPhC.133..128S}. The code performs simultaneous expansion of time-independent coupled-channel equations in Delves hyperspherical coordinates for all three possible chemical arrangements of the products (2$\times$H$_2$+O, OH+H) \cite{1987JChPh..87.3888P} and solves them using one of the implemented propagators. Thus, the overcomplete basis set was used to avoid the coordinate selection problem for different configurations. The calculation was performed independently for each value of the total angular momentum quantum number $J$ and the triatomic parity $P = \pm 1$. 
% The results are given as a parity-adapted $S$-matrix $S^{J,P}$ \cite{2000CoPhC.133..128S}. 
State-to-state partial cross sections for the collision energy $E$, describing transitions from the initial ro-vibrational state $(v,j)$ of H$_2$ molecule to final quantum states $(v'',j'')$ of H$_2$ or to states $(v',j')$ of OH molecule, are constructed from the helicity representation of the $S$-matrix according to \cite{2000CoPhC.133..128S}
\begin{equation}
 \sigma_{vj \rightarrow v'j'}^{J}(E) = \frac{\pi}{k_{vj}^2} \frac{2 J + 1}{2 j + 1} \sum_{v'j'k'k} |S^J_{vjk, v'j'k'} \left(E \right) |^2 ,
 \label{eq2}
\end{equation}
where, for the incident channel $(v,j)$, the wave vector is denoted as $k_{vj}$ and corresponding $S$-matrix elements as $S^J_{vjk, v'j'k'}(E)$. These expressions are valid both for the complete and a restricted basis set, where the angular momentum quantum numbers $k$ and $k'$ are defined as $0 \leqslant k \leqslant \min(J,j)$ and $0 \leqslant k' \leqslant \min(J,j')$.

The observables are calculated from the integral reactive cross section defined as
\begin{equation}
 \sigma_{vj \rightarrow v'j'}(E) = \sum_{J=0}^{J_{\mathrm{max}}} \sigma_{vj \rightarrow v'j'}^J(E) .
 \label{eq1}
\end{equation}
We also define the total reactive cross section $\sigma_{vj}$ for the initial state H$_{2}(v,j)$ as 
\begin{equation}
  \sigma_{vj}(E) = \sum_{v'j'} \sigma_{vj \rightarrow v'j'}(E) ,
\end{equation}
where the summation is performed over all energeticaly open states OH$(v',j')$.

The state-to-state momentum transfer cross section for the initial state $(v,j)$ and final state $(v',j')$, of interest in studies of transport properties, is given by \cite{1960RSPSA.256..540A,1962PApPh..13..643D}
\begin{equation}
 \sigma^{\mathrm{mt}}_{vj \rightarrow v'j'}(E) = 2\pi \int_{0}^{\pi} Q_{vj,v'j'}(\theta,E) \sin \theta (1 - \cos \theta) \mathrm{d} \theta, 
 \label{eq3}
\end{equation}
where $Q_{vj,v'j'}(\theta,E) = {d \sigma_{vj \rightarrow v'j'}(\theta,E)}/{d \Omega}$ is the differential cross section given by\cite{2000CoPhC.133..128S}
\begin{equation}
%  Q_{vj,v'j'}(E,\theta) = \frac{\pi}{k^2_{vj}} \sum_J (2J+1)|S^J_{vjk,v'j'k'}(E)|^2 .
 Q_{vj,v'j'}(\theta,E) = \left\lvert \frac{1}{2 i k_{vj}} \sum_J (2J+1) d_{k'k}^{J}(\theta) S^J_{vjk,v'j'k'}(E) \right\rvert^2 .
\end{equation}
State-to-state partial rate coefficients are defined by averaging the cross sections over Maxwell-Boltzmann distributions of the colliding particles' relative velocities:
\begin{equation}
  k_{vj,v'j'}(T) = \sqrt{\frac{8}{\pi \mu (k_B T)^3}} \int_0^\infty{\sigma_{vj \rightarrow v'j'}(E)} e^{-\frac{E}{k_B T}} E dE .
\end{equation}
% For comparing these quantities with experimental results it is useful to find their sums over all energeticaly allowed states  $(v',j')$ of the product molecule for a selected initial state H$_{2}(v,j)$. 
A major difficulty associated with the O($^{3}P$) + H$_2$ reaction is the presence of coupled, multiply-degenerate potential energy surfaces\cite{2003JChPh.11912360M}. While a full model of the reaction should include spin-orbit and nonadiabatic coupling terms between all surfaces, several authors have demonstrated\cite{2003JChPh.11912360M,2005JChPh.122u4301C} that these terms can be neglected if the collision energy is significantly larger than the coupling strength.
Specifically, a recent QCT study\cite{2005JChPh.122u4301C} of surface-hopping between O($^{3}P$) + H$_2$ and O($^{1}D$) + H$_2$ reactions concluded that the reaction can be described by a single surface for the considered collision energy interval. The fact that the cross sections obtained from crossed molecular beam experiments\cite{2003JChPh.118.1585G} agree well with the calculations performed on uncoupled surfaces \cite{2004JChPh.121.6346B} further supports this argument. 
Thus, we carried out independent single-surface calculations on the Rogers' GLDP\cite{Rogers_Kupperman_PES_2000} $^{3}A'$ and Branda\~o's BMS1\cite{2004JChPh.121.8861B_PES} $^{3}A''$ PESs. 
The total cross sections are given as the sum of the two calculations weighted by a multiplicity factor of 1/3\cite{1981JChPh..74.4984S}. We expect this approximation to be valid at collision energies greater than the asymptotic spin-orbit splitting of the $^3P$ term of the oxygen atom, which is about 0.02 eV.

\subsection{Calculation details and convergence tests}

State-to-state cross sections for the H$_2(v,j)$ + O$(^{3}P)$ reaction were calculated using the ABC code\cite{2000CoPhC.133..128S}. Balakrishnan \cite{2003JChPh.119..195B,2004JChPh.121.6346B} used a similar approach to calculate the reactive cross sections for this reaction for incident kinetic energies up to 1.2 eV. Extending the calculation up to the collision energy of 4.4 eV, of interest in atmospheric science and astrophysics, required a significantly larger basis set to correctly describe the population of the excitated ro-vibrational
states. The relevant parameters that control the truncation of the basis set \cite{2000CoPhC.133..128S} were: the cutoff energy $E_{\mathrm{max}}$, the maximum allowed rotational level $j_{\mathrm{max}}$, the maximum value of the angular momentum projection quantum number $k_{\mathrm{max}}$, and the total angular momentum quantum number $J$. 
\begin{figure}[t]
\includegraphics[width=20pc]{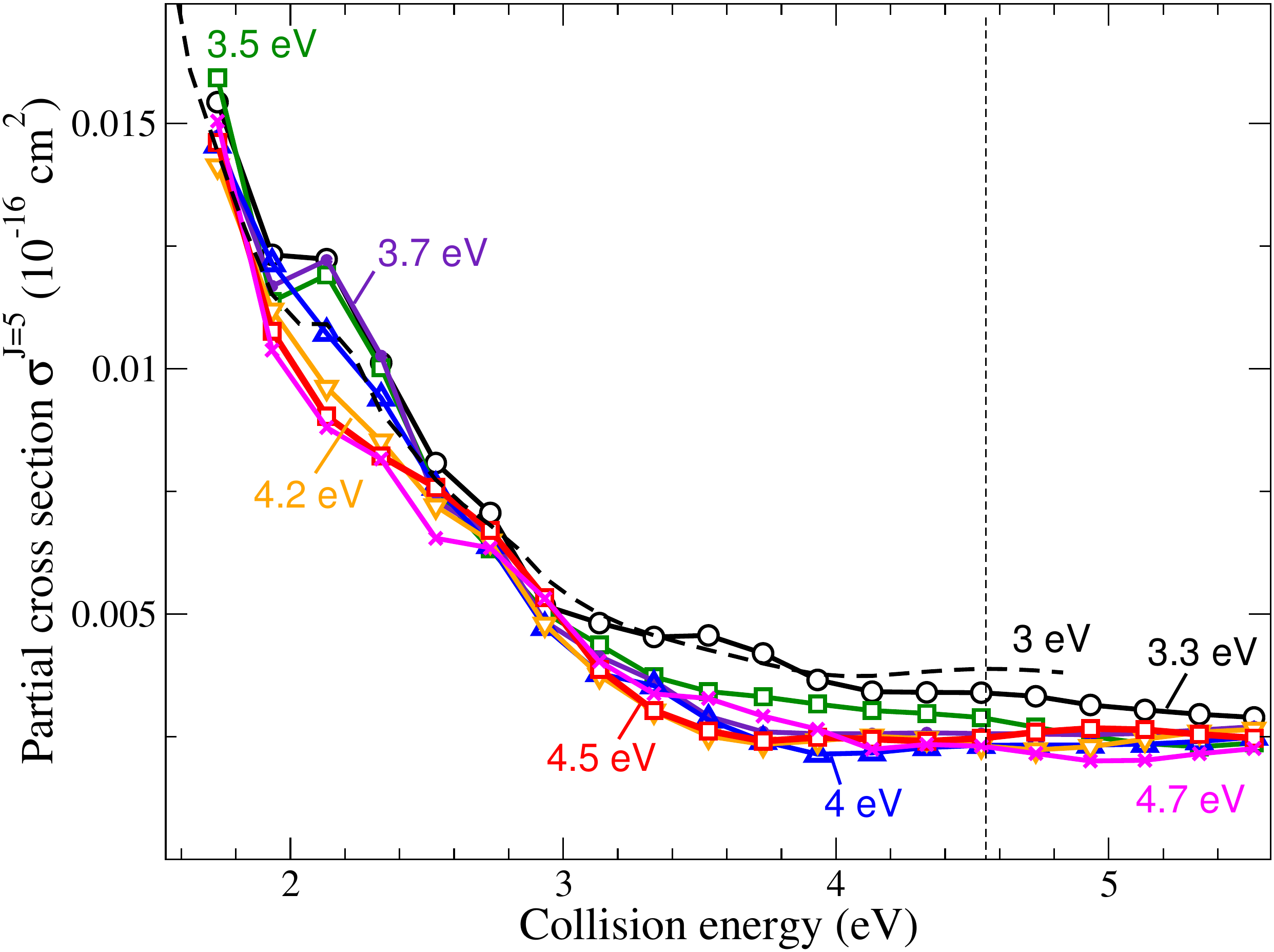}
\caption{\label{fig1} Convergence of the partial cross section $\sigma^{J=5}_{00 \rightarrow v'j'}$ on the $^3A''$ PES with respect to the cutoff energy $E_{\mathrm{max}}$. The parameters $j_{\mathrm{max}}=18$ and $k_{\mathrm{max}}=4$ are used. For $E \leqslant 1.7$ eV the convergence for all $E_{\mathrm{max}}$ was satisfactory. Vertical dashed line at $E_{\mathrm{max}}= 4.5$ eV indicates the upper limit of reliable convergence.}
\end{figure}
\begin{figure}[t]
\includegraphics[width=20pc]{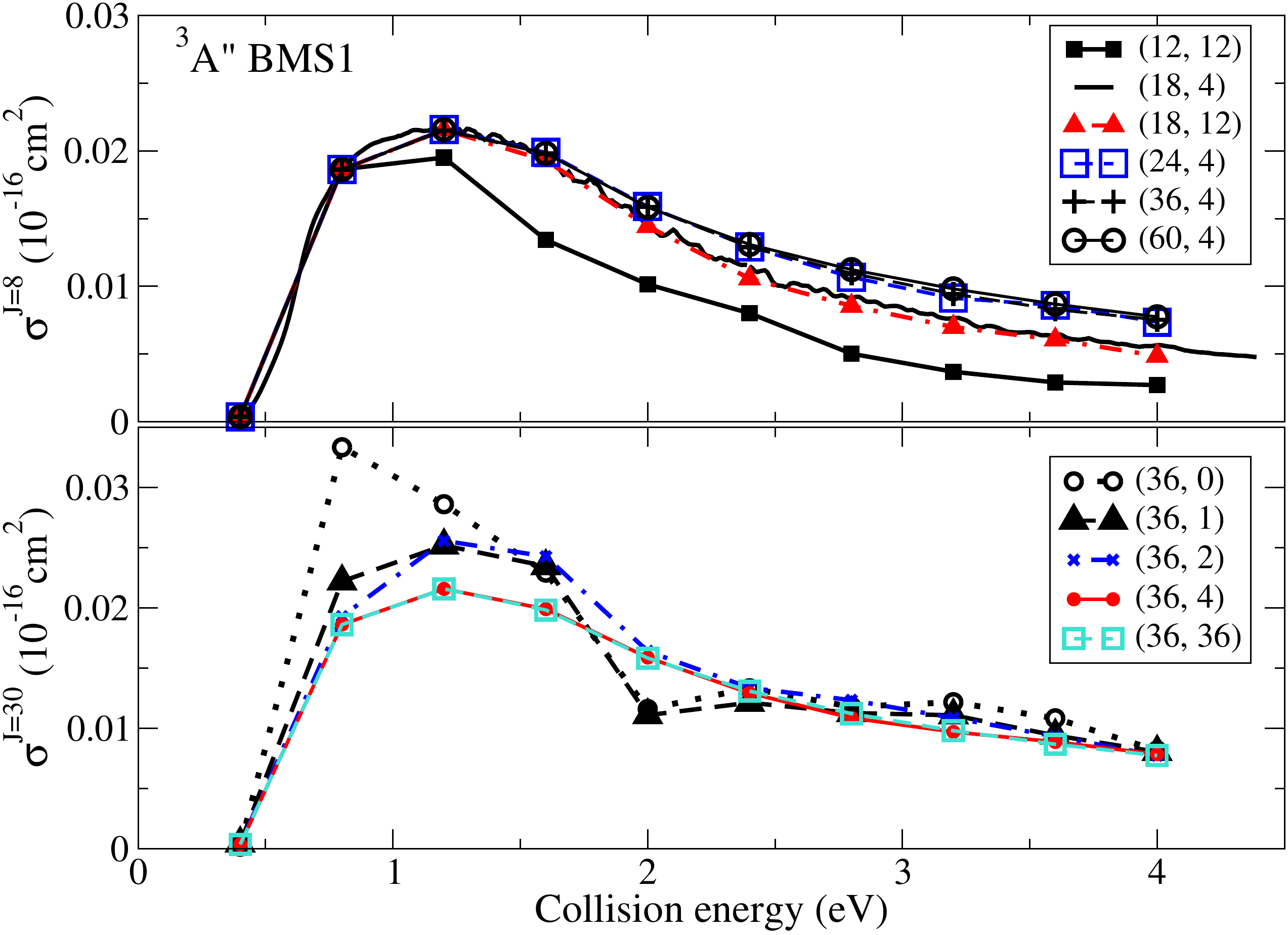}
\caption{\label{fig2}Partial cross section $\sigma^{J=30}_{00 \rightarrow v'j'}$ calculated on the $^3A''$ PES for a number of $(j_{\mathrm{max}},k_{\mathrm{max}})$ basis truncation parameters (upper panel). The convergence with respect to $k_{\mathrm{max}}$ for $j_{\mathrm{max}}=36$ is given in the lower panel. 
The production run parameters $(j_{\mathrm{max}}=36, k_{\mathrm{max}}=4)$ are indicated by thick red line.}
\end{figure}
Extensive convergence tests on both PESs were performed. The convergence of the numerical differential equation solver in the energy range of interest was achieved for the propagation hyperradius $\rho_{\mathrm{max}}=25.0$ $a_0$ and step size $\Delta\rho = 0.02$ $a_0$, in agreement with the related studies\cite{2004JChPh.121.6346B,2012GeoRL..3910203G}.
The required cutoff energy $E_{\mathrm{max}}$ was determined by calculating the partial cross sections for the range of cutoff energies from 3 to 4.7 eV for the selected values of $J$, $j_{\mathrm{max}}$, and $k_{\mathrm{max}}$ (Fig. \ref{fig1}). Regardless of the total angular momentum $J$, we observed little improvement in convergence for the cutoff energy greater than $E_{\mathrm{max}}=4.3$ eV, or from increasing $j_{\mathrm{max}}$ and $k_{\mathrm{max}}$ past the illustrated values. In fact, keeping the cutoff energy as low as 3 eV resulted in a surprisingly accurate calculation even at the collision energies as high as 4.5 eV. 
This can be understood from the fact that while the density of states is greatly increased close to the dissociation energies of OH+H and O+H$_2$, reported to be at 4.58 and 4.7 eV \cite{Rogers_Kupperman_PES_2000}, the impact of the added channels to the population distribution in the reaction is rather small.
Note that for $E_{\mathrm{max}}>4.5$ eV the basis size grows rapidly and makes the calculation prohibitively expensive.

The convergence of partial cross sections with respect to $j_{\mathrm{max}}$ and $k_{\mathrm{max}}$ on the BMS1 $^3A''$ PES, for which the calculation was slower to converge, is illustrated in Fig. \ref{fig2} for a low and a high value of $J$. 
The parameter space defined by the $(j_{\mathrm{max}}, k_{\mathrm{max}})$ pair, where $j_{\mathrm{max}}=12 \ldots 60$ and $k_{\mathrm{max}}=0 \ldots 36$, was explored for the energy cutoff set to $E_{\mathrm{max}}=4.5$ eV.
The convergence was achieved for $j_{\mathrm{max}} \geqslant 36$ and $k_{\mathrm{max}} \geqslant 12$. Restricting the maximum projection quantum number to $4 \leqslant k_{\mathrm{max}} < 12$ was found to reduce the calculation time by a factor of 4 to 10 without a significant loss of precision (see Fig. \ref{fig2}, bottom panel). 
Based on the convergence analysis, the production runs were carried out using $E_{\mathrm{max}}=4.5$ eV, $j_{\mathrm{max}} \geqslant 36$, and $k_{\mathrm{max}} = 6$, resulting in 3,266 and 3,300 coupled channels for $^{3}A'$ and $^{3}A''$ PES, respectively.

Finally, we tested the convergence with respect to the total number of partial waves included in the calculation by directly comparing their contributions to the total cross section. It took a total of $J_{\mathrm{max}}=105$ partial waves to achieve the convergence of reactive and inelastic cross sections to within a few percent for the highest collision energy considered. 
The complete production run took about 8,470 CPU-days, or about 18.2 hours per an energy point, per a single core of Intel Xeon$^\circledR$ X5670 CPU.

\section{Results and discussion}

\subsection{O($^{3}P$) + H$_2(v=0,j=0) \rightarrow$ H + OH$(v',j')$}

\begin{figure}[t]
\includegraphics[width=20pc]{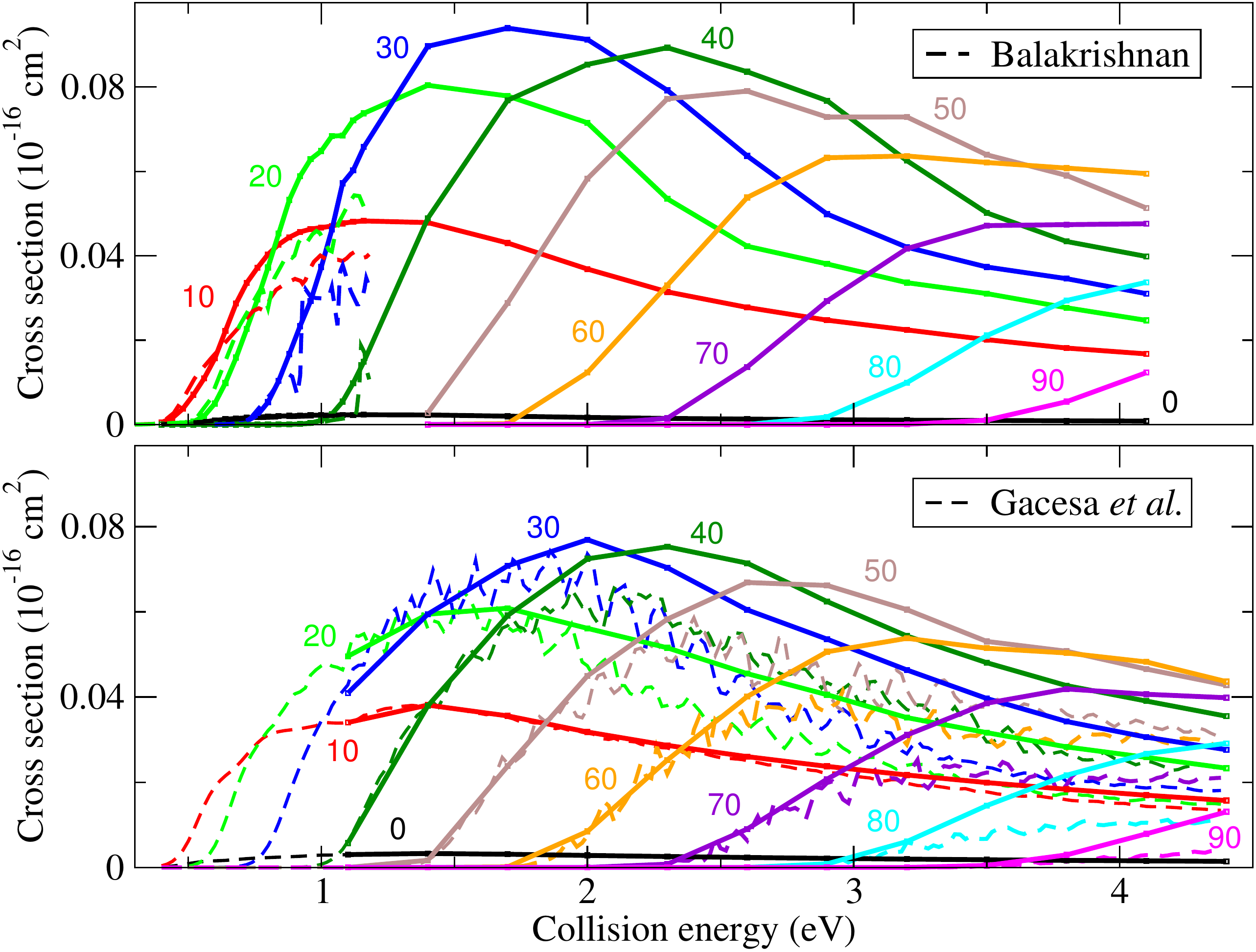}
\caption{\label{fig3} Partial cross sections $\sigma^{J}_{00}$ summed over all target states in dependence of the incident kinetic energy for the O$(^3P)$+H$_2(v=0,j=0) \rightarrow$ H + OH$(v',j')$ reaction, shown for selected values of the total angular momentum quantum number $J$ (numbers). \textit{Top:} Results for $^{3}A''$ PES are compared with the calculation by Balakrishnan\cite{2004JChPh.121.6346B} (dashed lines). \textit{Bottom:} Results for the $^{3}A'$ PES. A calculation by Gacesa \textit{et al.}\cite{2012GeoRL..3910203G} is also shown (dashed lines).}
\end{figure}

We report results for O($^{3}P$) + H$_2(v=0,j=0)$ reactive scattering on both PESs considered. Partial reactive cross sections, defined as $\sigma^{J}_{vj}(E) = \sum_{v'j'} \sigma^{J}_{vj \rightarrow v'j'}(E)$, are compared to the results reported by Balakrishnan\cite{2004JChPh.121.6346B} on Rogers' GLDP $^{3}A''$ PESs (Fig. \ref{fig3}, top panel). The agreement between the two sets of results is good, although our cross sections remain consistently larger with the more pronounced difference at higher collision energies. The exception is $J=0$ partial cross section, where the agreement is perfect. In case of the $^{3}A'$ PES, we compare the partial cross sections to our earlier calculation\cite{2012GeoRL..3910203G} (Fig. \ref{fig3}, bottom panel). 
We believe that a larger basis set used in the current study is able to better account for higher rotational states populated during the reaction, yielding somewhat larger cross sections. 
Note that the oscillations present for higher partial waves in the previous high resolution calculation\cite{2012GeoRL..3910203G} are due to the centrifugal barrier of the rotating atom-molecule composite.

\begin{figure}[t]
\includegraphics[width=20pc]{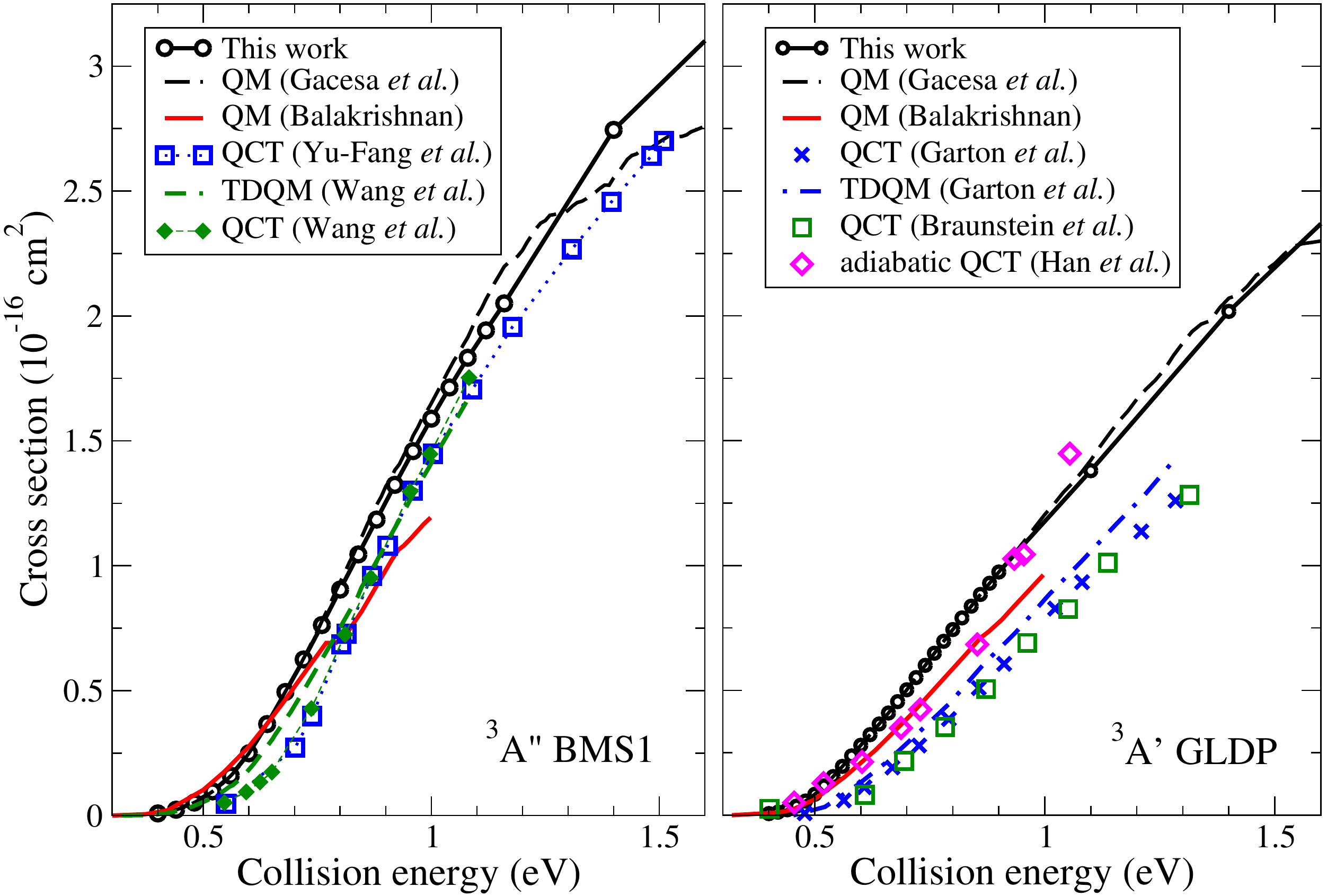}
\caption{\label{fig4} Integral reactive cross sections for O$(^3P)$+H$_2(v=0,j=0)$ for the BMS1\cite{2004JChPh.121.8861B_PES} $^3A''$ (left panel) and GLDP\cite{Rogers_Kupperman_PES_2000} $^3A'$ (right panel) PES as a function of the collision energy up to 1.6 eV.
}
\end{figure}
The reactive cross sections for O$(^3P)$+H$_2(v=0,j=0) \rightarrow $ H + OH($v',j'$) were calculated separately on each PES according to Eq. (\ref{eq1}), where the summation was performed over the partial cross sections up to $J_\mathrm{max}=105$. A comparison with selected published results\cite{2004JChPh.121.6346B} at lower collision energies is given in Fig. \ref{fig4}. 
Our cross sections are larger than the existing QM and quasi-classical trajectory (QCT) results for both PESs, except at low energies, where the agreement is good. The only exception is the adiabatic QCT calculation by Han \textit{et al.}\cite{JCC:JCC21940} at $E\gtrapprox0.9$ eV, which predicted comparable or higher values.
In case of the $^3A''$ PES, our results match previous QM calculations\cite{2003JChPh.119..195B,2004JChPh.121.6346B} below 0.7 eV, but deviate at higher energies, with the difference increasing up to about 15 \% at 1 eV. However, a direct comparison is not possible since these studies used Rogers' $^3A''$ PES\cite{Rogers_Kupperman_PES_2000}.
Note that the inflection feature, present in Balakrishnan\cite{2004JChPh.121.6346B} at about 0.8 eV, was not reproduced by our calculation. 
The agreement at high energies was better with recent QCT and TDQM results of Wang \textit{et al.}\cite{2006CPL...418..250W} and Yu-Fang \textit{et al.}\cite{1674-1056-20-7-078201}, where the cross sections agree qualitatively, although our results remain larger by about $(0.1-0.15) \times 10^{-16}$ cm$^2$. The best overall agreement is with the TDQM results of Wang \textit{et al.}\cite{2006CPL...418..250W}. 
A similar behavior is present for the $^3A'$ PES, except in case of the adiabatic QCT calculation by Han \textit{et al.}\cite{JCC:JCC21940}, which predicted comparable or higher cross sections for $E>0.9$ eV. Other QM studies on the BMS1 surface\cite{2006JChPh.124g4308W,2006JPhB...39.1215W} were conducted at lower energies, making direct comparison with our results impossible.

The differences between our results and the QM result of Balakrishnan\cite{2004JChPh.121.6346B} can be explained by the fact that the cross sections converge better at higher energies for a larger basis set, as can be seen from the convergence of partial cross sections (Fig. \ref{fig2}). The fact that our cross sections are larger for the $^3A'$ surface as well implies that this argument is independent of the PES.
Nevertheless, we cannot rule out the possibility that the differences in the van der Waals region\cite{2004JChPh.121.8861B_PES,B608871F} between the Rogers' GLDP\cite{Rogers_Kupperman_PES_2000} and Branda\~o's refitted BMS1\cite{2004JChPh.121.8861B_PES} $^3A''$ PESs, or the absence of nonadiabatic effects\cite{JCC:JCC21940}, are sufficient to affect the cross sections resulting in a better agreement at low energies.

\begin{figure}[t]
\includegraphics[width=20pc]{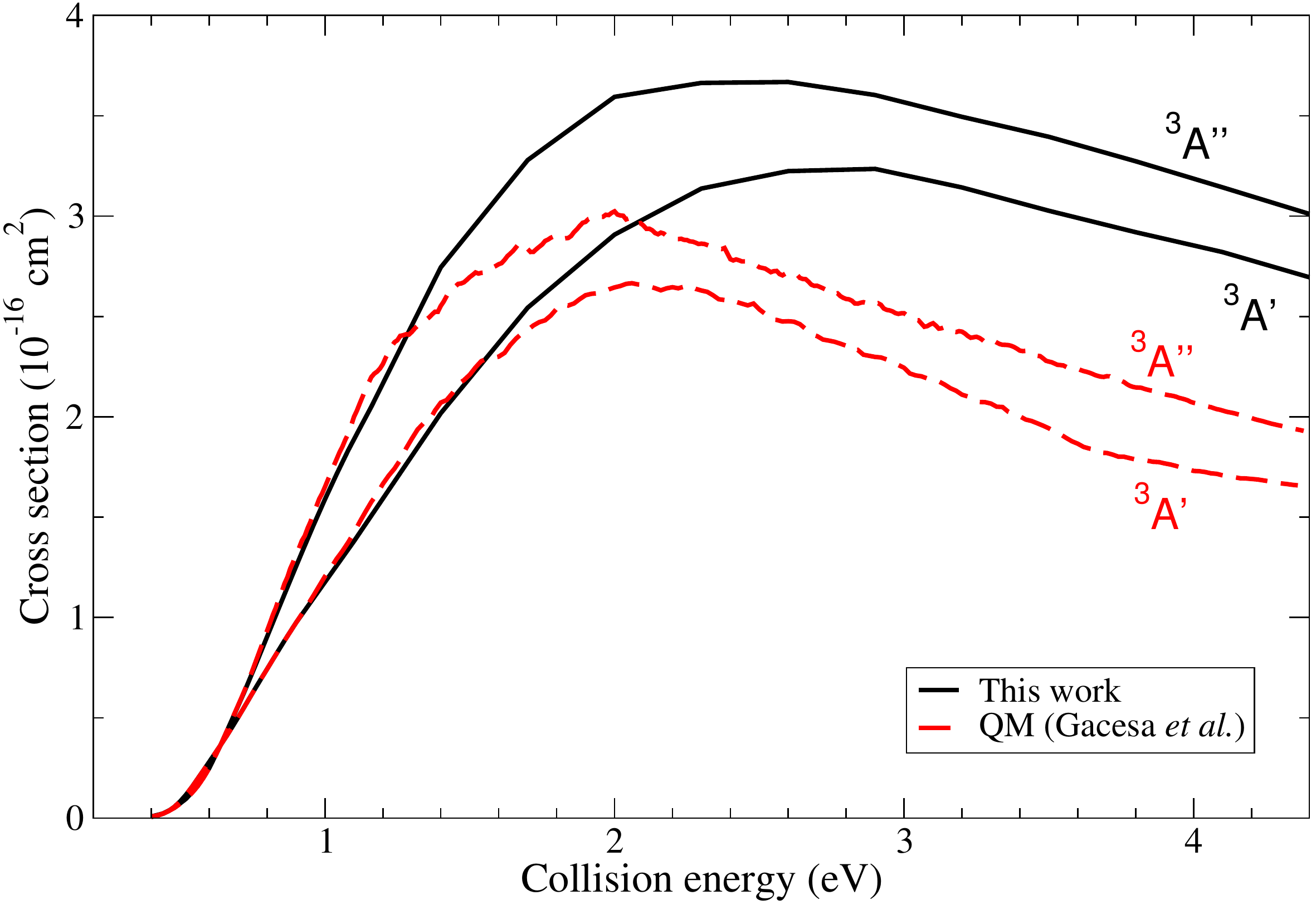}
\caption{\label{fig5} Integral reactive cross sections for O$(^3P)$+H$_2(v=0,j=0)$ reaction given for collision energies from 0.4 to 4.4 eV.}
\end{figure}
Reactive cross sections calculated for the complete energy range from 0.4 to 4.4 eV are given in Fig. \ref{fig5}, and compared with our earlier calculation\cite{2012GeoRL..3910203G}. 
Both sets of curves exhibit qualitatively similar behavior and remain in excellent agreement up to about 1.4 eV. At higher energies, the basis set of Gacesa \textit{et al.}\cite{2012GeoRL..3910203G} does not fully capture the reaction dynamics and predicts lower population distributions in the higher rotational states, resulting in smaller cross sections. The results from Fig. \ref{fig5} are given in Table \ref{table1}.

\begin{figure}[b]
\includegraphics[width=20pc]{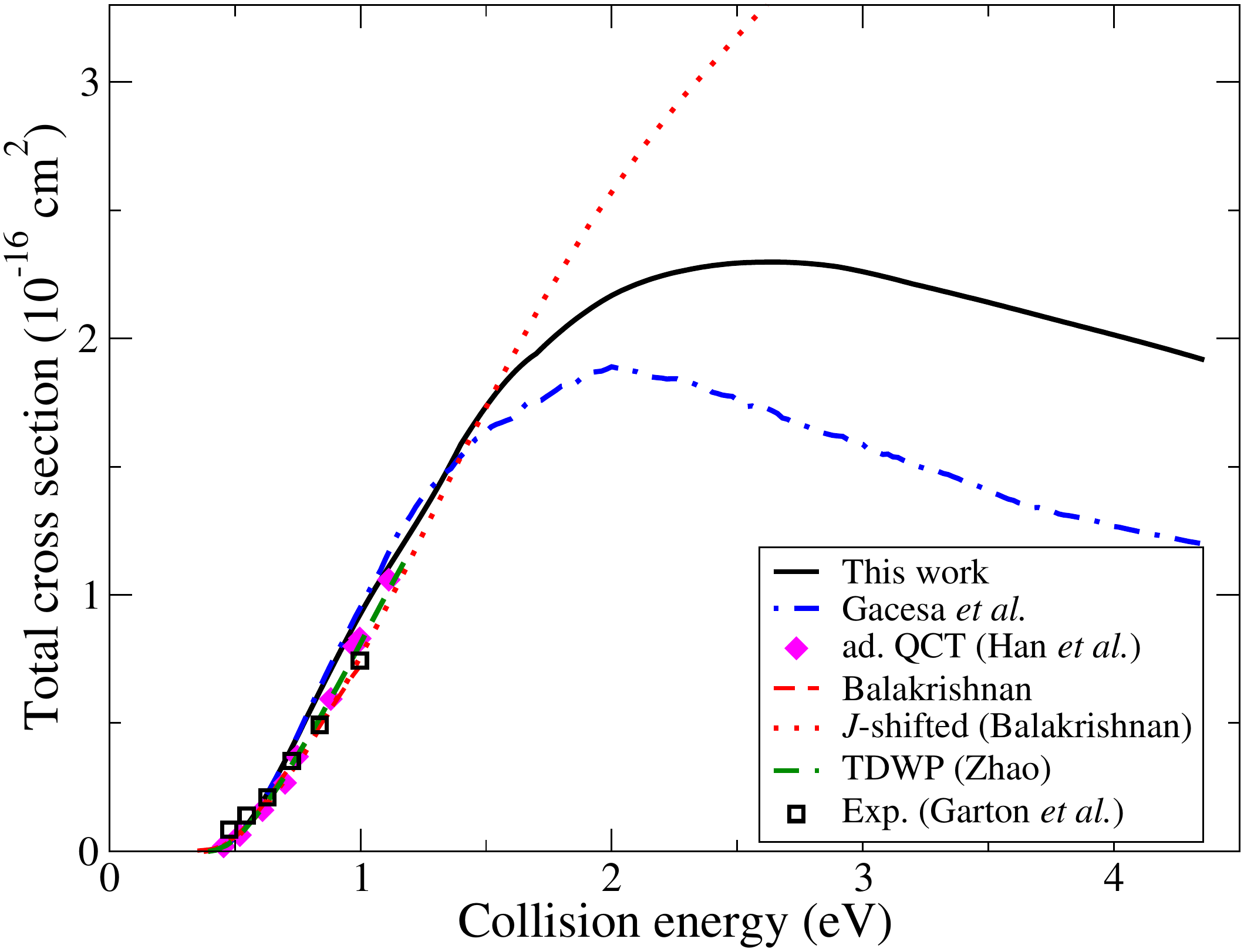}
\caption{\label{fig6} Total cross section for O$(^3P)$+H$_2(v=0,j=0)$ reaction as a function of the collision energy. Recent theoretical results by Balakrishnan\cite{2004JChPh.121.6346B}, Gacesa \textit{et al.}\cite{2012GeoRL..3910203G}, and Zhao\cite{10.1063/1.4795497}, and experiment by Garton \textit{et al.}\cite{2003JChPh.118.1585G} are also shown.
}
\end{figure}

\begin{table}[tb]
\caption{\label{table1}Integral cross sections for O$(^3P)$+H$_2(v=0,j=0)$ reaction as a function of the collision energy calculated on $^3A'$ and $^3A''$ PESs.}
\begin{ruledtabular}
\begin{tabular}{cccc}
           & \multicolumn{3}{c}{$\sigma_{v=0,j=0}$ $(10^{-16}$ cm$^2$)} \\
   $E$(eV) & $^3A'$ & $^3A''$ & $\left(^3A'+^3A''\right)/3$ \\
   \hline
      0.40 & $6.62 \times 10^{-3}$ & $8.31 \times 10^{-3}$ & $4.98 \times 10^{-3}$ \\
      0.44 & $2.18 \times 10^{-2}$ & $2.31 \times 10^{-2}$ & $1.50 \times 10^{-2}$ \\
      0.48 & $5.69 \times 10^{-2}$ & $5.20 \times 10^{-2}$ & $3.63 \times 10^{-2}$ \\
      0.52 & 0.12 & $9.65 \times 10^{-2}$ & $7.13 \times 10^{-2}$ \\
      0.56 & 0.20 & 0.16 & 0.12 \\
      0.60 & 0.28 & 0.25 & 0.18 \\
      0.64 & 0.37 & 0.37 & 0.24 \\
      0.68 & 0.46 & 0.49 & 0.32 \\
      0.72 & 0.55 & 0.63 & 0.39 \\
      0.76 & 0.65 & 0.76 & 0.47 \\
      0.80 & 0.74 & 0.91 & 0.55 \\
      0.84 & 0.84 & 1.05 & 0.63 \\
      0.88 & 0.93 & 1.18 & 0.70 \\
      0.92 & 0.99 & 1.32 & 0.78 \\
      1.00 & 1.16 & 1.59 & 0.93 \\
      1.10 & 1.38 & 1.87 & 1.09 \\
      1.40 & 2.02 & 2.74 & 1.59 \\
      1.70 & 2.54 & 3.28 & 1.94 \\
      2.00 & 2.91 & 3.59 & 2.17 \\
      2.30 & 3.14 & 3.66 & 2.27 \\
      2.60 & 3.22 & 3.67 & 2.30 \\
      2.90 & 3.24 & 3.60 & 2.28 \\
      3.20 & 3.14 & 3.49 & 2.21 \\
      3.50 & 3.03 & 3.40 & 2.14 \\
      3.80 & 2.92 & 3.27 & 2.06 \\
      4.10 & 2.82 & 3.14 & 1.99 \\
      4.40 & 2.69 & 3.01 & 1.90 \\
\end{tabular}
\end{ruledtabular}
\end{table}

The total cross section for the O$(^3P)$+H$_2(v=0,j=0)$ reaction was calculated as the sum $\sigma_{vj}(E) = \left( \sigma_{vj}^{(^{3}A'')}(E) + \sigma_{vj}^{(^{3}A')}(E) \right) / 3$, as described above, and shown in Fig. \ref{fig6}. The QM results of Balakrishnan extended to higher energies\cite{2004JChPh.121.6346B,2005ApJ...629..305S} by means of $J$-shifting approximation\cite{doi:10.1021/j100166a014}, time-dependent wave packet (TDWP) calculation of Zhao\cite{10.1063/1.4795497}, and the experimental results of Garton \textit{et al.}\cite{2003JChPh.118.1585G} are given for comparison. A notable difference from the calculation performed using $J$-shifting approximation at higher collision energies warrants further explanation.

The $J$-shifted total cross section is constructed using Eq. (\ref{eq1}), where the initial state-selected reaction probability $P_{vj}^{J}$ is calculated exactly for $J=0$ and approximated for $J>0$ by shifting the collision energy according to\cite{doi:10.1021/j100166a014,2004JChPh.121.6346B}
\begin{equation}
  P_{vj}^J \approx P_{vj}^{J=0} (E - E_{J}^{\ddag}) ,
\end{equation}
where $E_{J}^{\ddag} = B^{\ddag} J(J+1)$ is the rotational energy of the linear triatomic transition state assumed by the complex during the reaction. The rotational constant $B^{\ddag}$ is taken to be 3.127 and 3.154 cm$^{-1}$ for the $^3A'$ and $^3A''$ PES, respectively\cite{2005ApJ...629..305S}. By repeating these steps we were able to closely match the result of Sultanov \textit{et al.}\cite{2005ApJ...629..305S} (Fig. \ref{fig6}, dashed curve). 
The $J$-shifting approximation has been shown to be accurate at low collision energies, up to about 1.2 eV\cite{doi:10.1021/j100166a014,2003JChPh.119..195B,2004JChPh.121.6346B,1987CPL...141..545B}. However, at higher energies the (OHH) transition state is no longer linear as the rotational energy is sufficiently high to overcome the reaction energy barrier. A study of the role of quantized transition states in the H$_2$+O reaction dynamics found that bending angles of the transition state complex up to 30\textdegree\, were present in the energy range up to 1.9 eV\cite{1993JChPh..98..342C}.

\begin{figure}[tb]
\includegraphics[width=20pc]{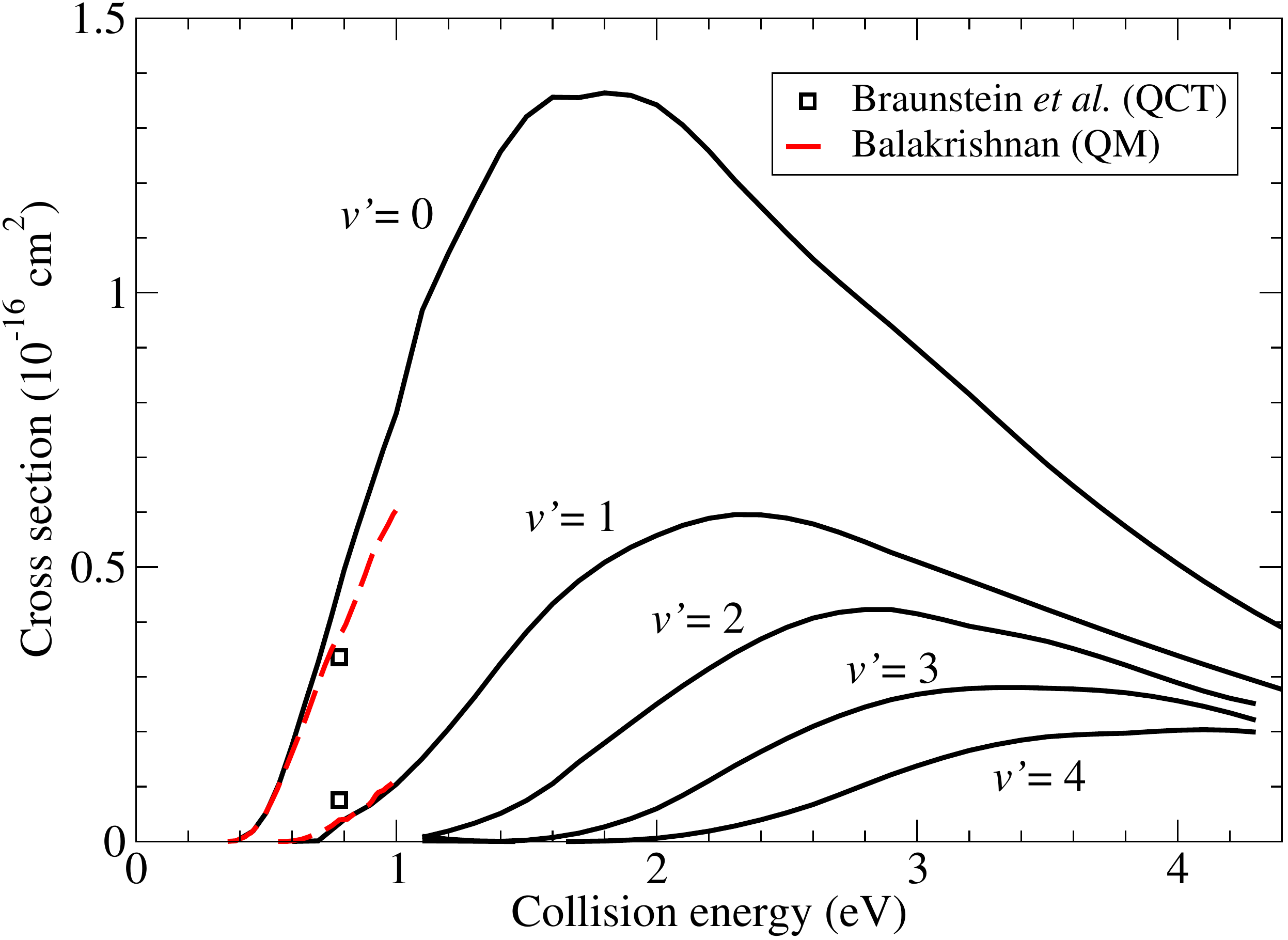}
\caption{\label{fig7} Cross sections for O$(^3P)$+H$_2(v=0,j=0) \rightarrow \mathrm{OH}(v') + \mathrm{H}$ reaction for $v'=0 \ldots 4$ as functions of collision energy. The QM results of Balakrishnan\cite{2004JChPh.121.6346B} and QCT results of Braunstein \textit{et al.}\cite{2004JChPh.120.4316B}, are shown.}
\end{figure}

At the collision energy greater than about 0.6 eV, the channels leading to production of vibrationally excited OH molecules become energeticaly allowed. In Fig. \ref{fig7} we show the cross sections in the considered energy range for the O$(^3P)$+H$_2(v=0,j=0) \rightarrow \mathrm{OH}(v') + \mathrm{H}$ reactive collision, where $v'=0 \ldots 4$. 
For high collision energies, $E>4.1$ eV, the vibrational states up to $v'=12$ can be populated. Our results are in good agreement with earlier QCT and QM results\cite{2004JChPh.121.6346B,2004JChPh.120.4316B} despite the fact that these were carried out on a different PES. The agreement with the QM result of Balakrishnan \textit{et al.}\cite{2004JChPh.121.6346B} for $v'=1$ is particularly good, possibly implying a better agreement between the PESs at the energies above 0.6 eV.

\subsection{Vibrational excitation of the reagent H$_2$}

\begin{figure}[tb]
\includegraphics[width=20pc]{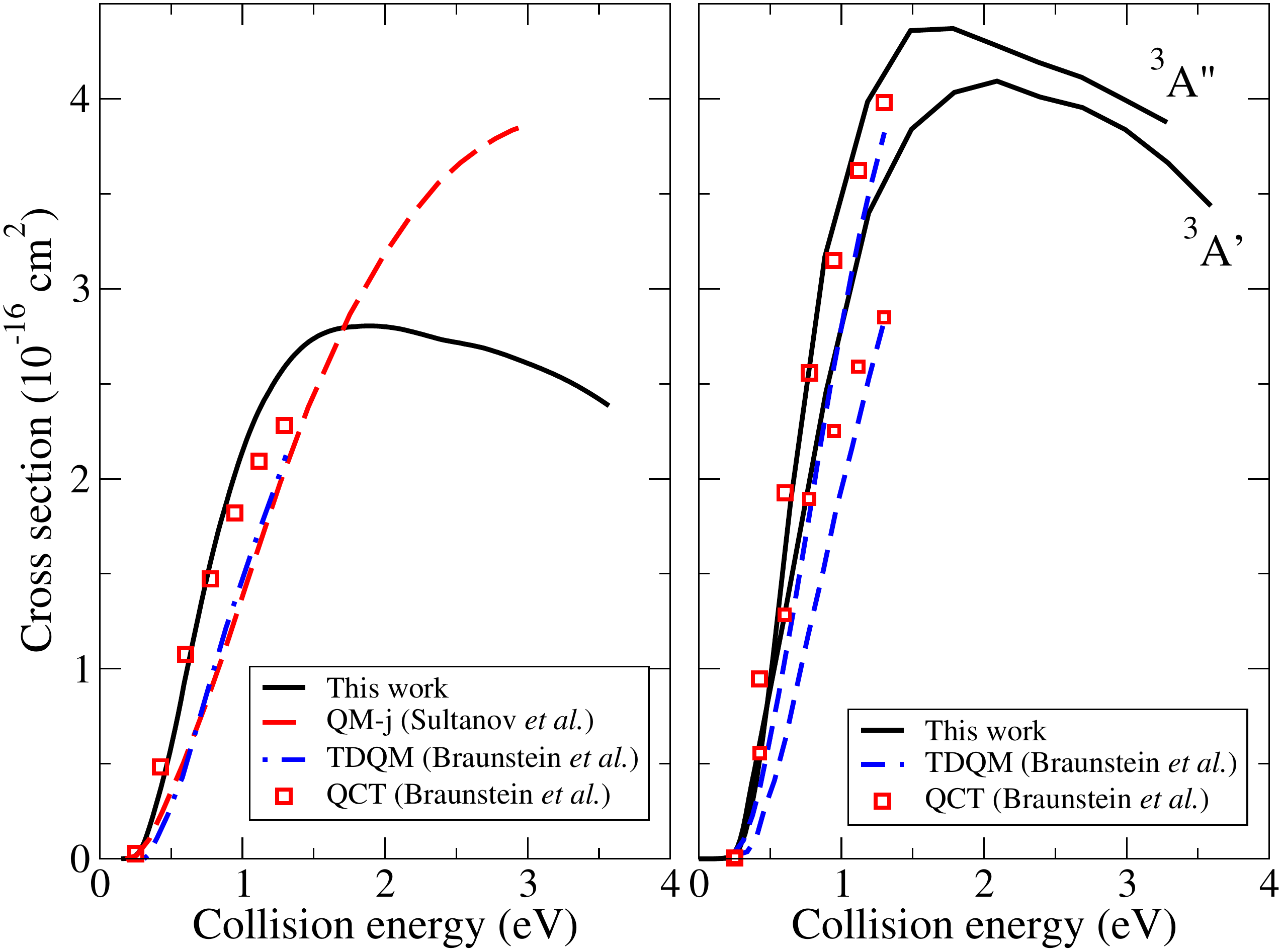}
\begin{flushleft}   
\caption{
\label{fig8} Total cross sections for O$(^3P)$+H$_2(v=1,j=0)$ reaction for statistically weighted (left panel) and individual surfaces (right panel). TDQM and QCT results of Braunstein \textit{et al.}\cite{2004JChPh.120.4316B} and $J$-shifted QM results of Sultanov \textit{et al.}\cite{2005ApJ...629..305S} are shown for comparison.
}
\end{flushleft}
\end{figure}

Fig. \ref{fig8} shows the present vibrationally-resolved reactive cross sections for the O$(^3P)$+H$_2(v=1) \rightarrow \mathrm{OH}(v') + \mathrm{H}$ reaction, both as a statistical sum over the $^3A'$ and $^3A''$ surfaces (left panel) and broken down into the individual contributions (right panel). 
We also show recent QCT and TDQM cross sections\cite{2004JChPh.120.4316B}, as well as QM results\cite{2003JChPh.119..195B,2004JChPh.121.6346B}, extended to intermediate energies, up to about 2.6 eV, via $J$-shifting approximation\cite{2005ApJ...629..305S}. 

As seen for the reaction with a non-vibrating hydrogen molecule, our results for the total cross section agree better with the QCT calculation (red squares) than with TDQM (blue dot-and-dash line) and QM (red dashed line) results. 
By examining individual contributions of the two surfaces (Fig. \ref{fig8}, right panel), we observe excellent agreement between the QCT results and our calculation for the $^3A''$ PES, while for the second surface the agreement remains very good until about 0.8 eV, where our cross section becomes increasingly larger as the collision energy increases. The TDQM and QM results remain consistently smaller for all but lowest energies. 

In their work, Braunstein \textit{et al.}\cite{2004JChPh.120.4316B} attribute the discrepancy between TDQM and QCT results to the lack of tunneling and other quantum effects in the classical description, and the differences in vibrational adiabaticity between classical and quantum mechanics, an effect previously studied in this system\cite{} at near-threshold energies.
While we believe this explanation is valid, it is difficult to estimate the relative importance of these effects versus the uncertainties introduced by the numerical methods, such as the cutoff in the basis set at higher energies, or, in case of $^3A''$, the differences in the PESs. The former could explain the disagreements in cross sections for the $^3A'$ surface. 
In case of the $^3A''$ surface, the cross sections were shown to be somewhat larger for near-threshold collisions\cite{2006JChPh.124g4308W} in $v=1$, as well as for non-vibrating reactants\cite{2006CPL...418..250W,1674-1056-20-7-078201,2012GeoRL..3910203G}. Since the QCT approach can result in larger cross sections than produced by quantum methods\cite{2004JChPh.120.4316B}, we cannot rule out the possibility that the excellent agreement with the QCT results is partly accidental. 

\begin{figure}[t]
\includegraphics[width=20pc]{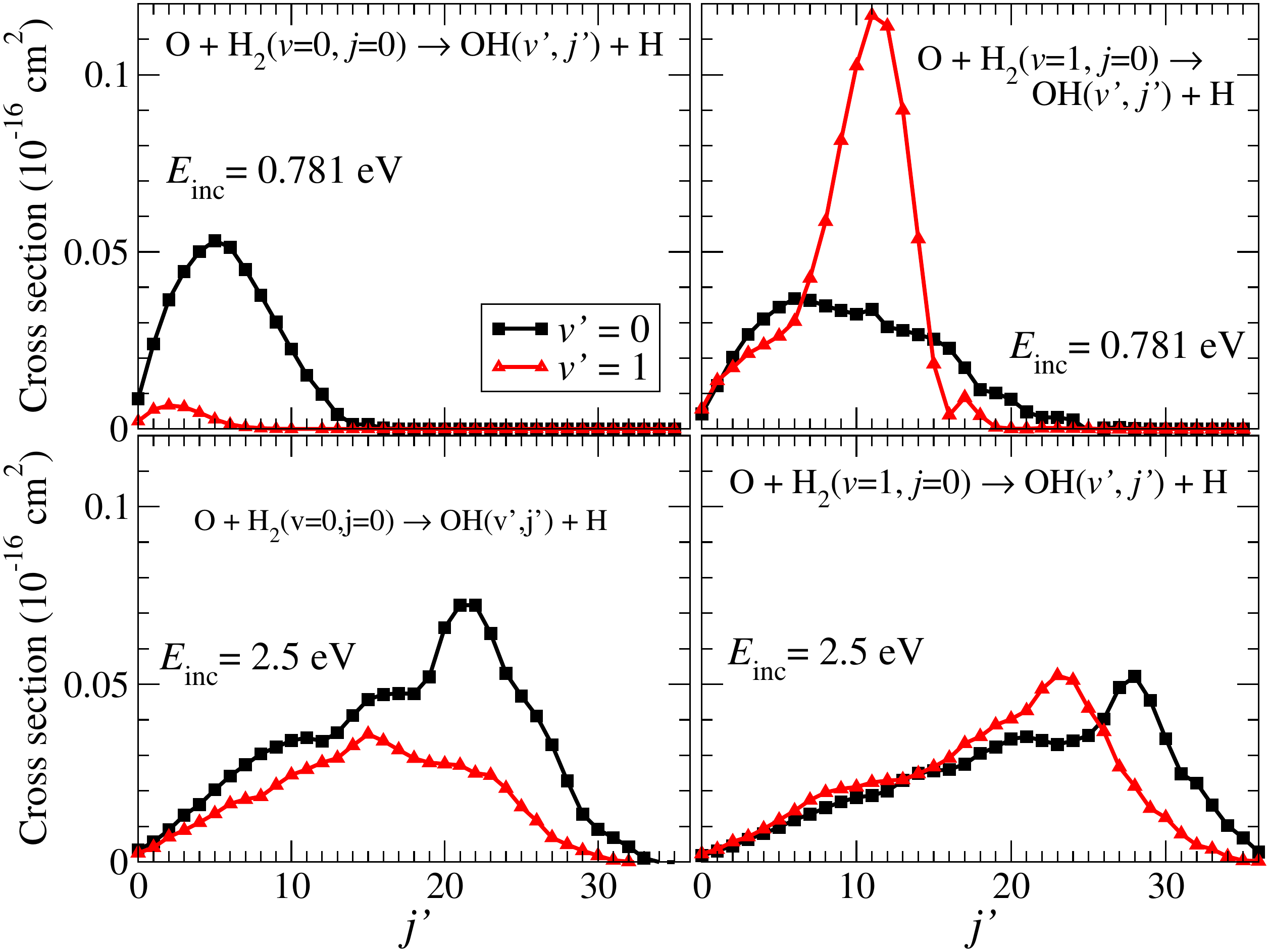}
\caption{\label{fig9} Rotationally resolved total cross sections for formation of OH$(v'=0-1,j')$ in O$(^3P)$+H$_2(v=0-1,j=0)$ reaction for the collision energy of 0.781 eV (top) and 2.5 eV (bottom).}
\end{figure}

\begin{figure*}[t]
\centering
   \begin{tabular}{@{}cc@{}}
      \includegraphics[width=.48\linewidth]{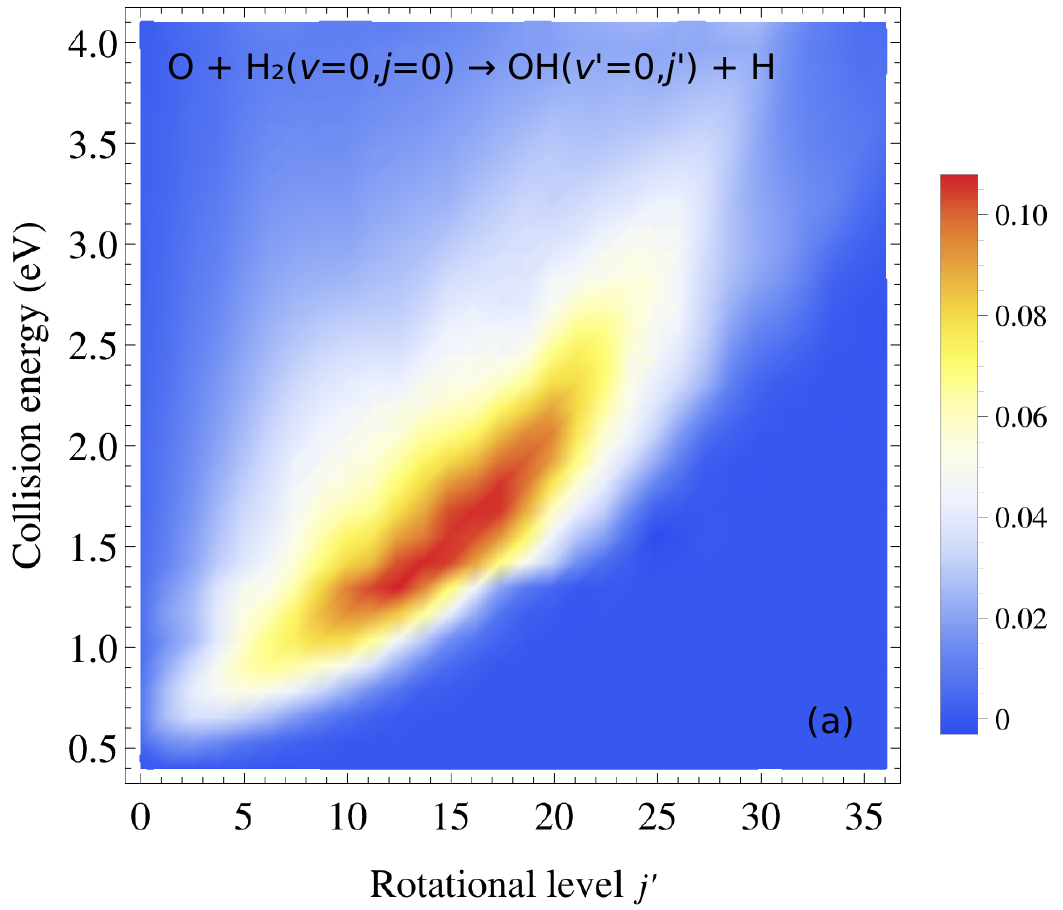} &
      \includegraphics[width=.48\linewidth]{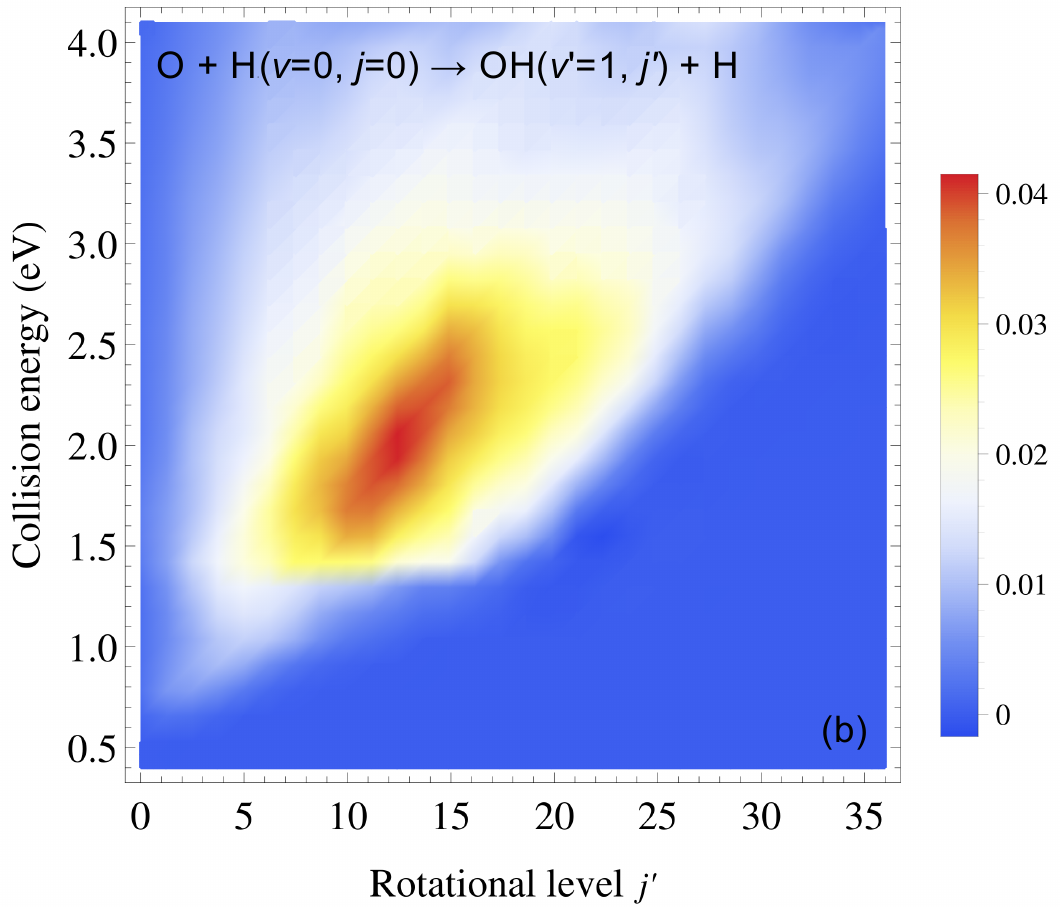} \\
      \includegraphics[width=.48\linewidth]{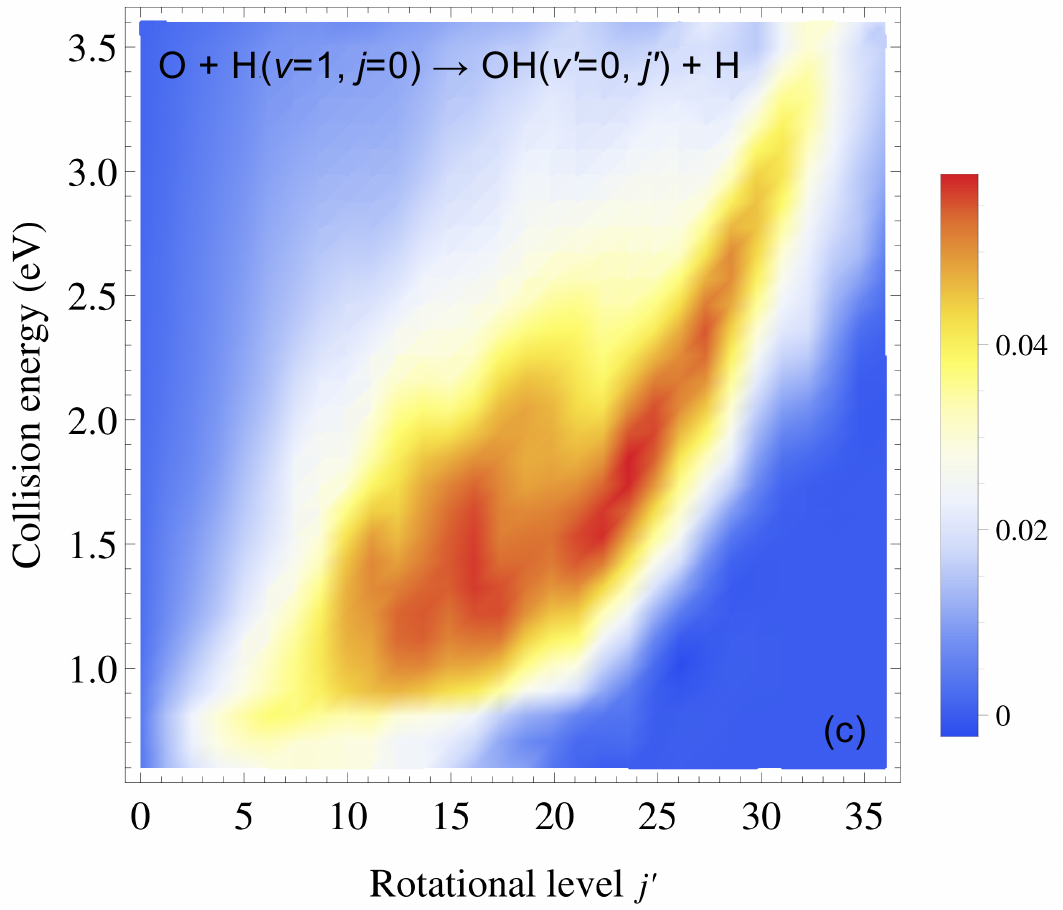} &
      \includegraphics[width=.48\linewidth]{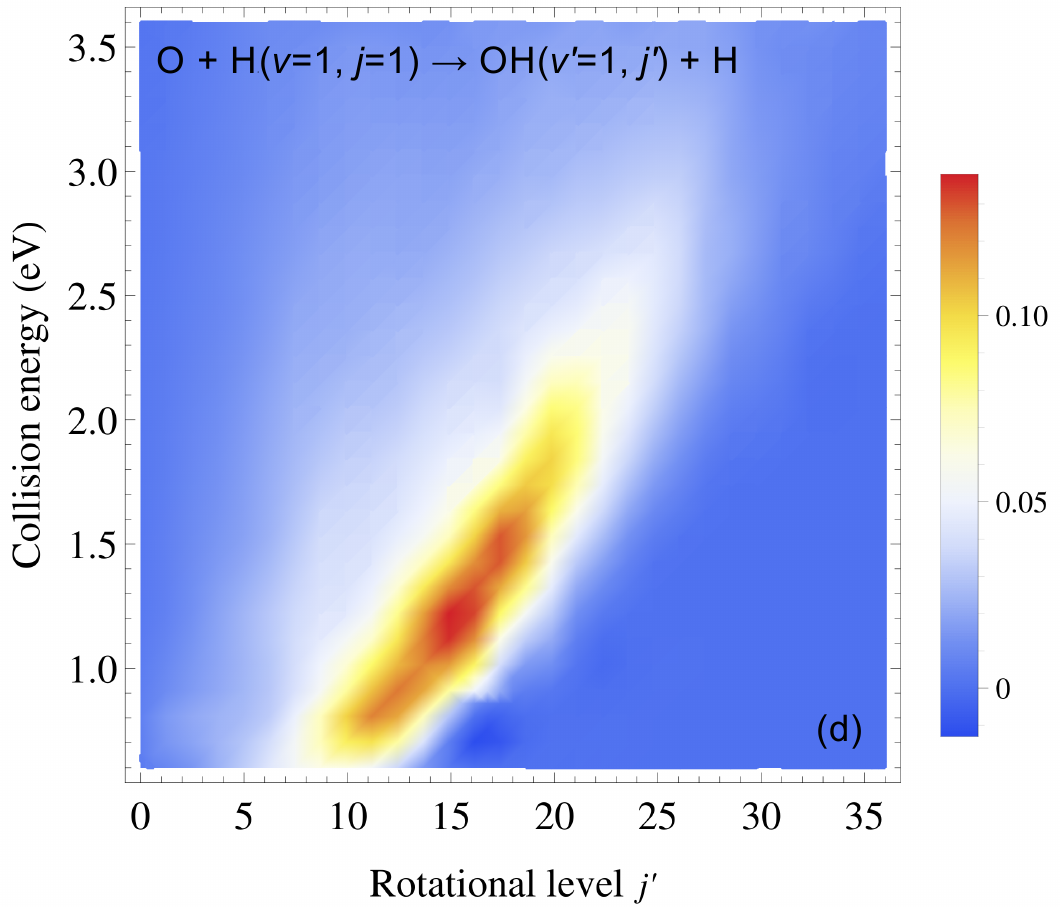} \\
   \end{tabular}
   \caption{\label{fig10} Rotationaly resolved total cross sections for O$(^3P)$+H$_2(v=0-1,j=0) \rightarrow $ OH($v'=0-1,j'$) + H reaction as a function of the collision energy.}
\end{figure*}

The statistically weighted total cross section peaks at the collisional energy of about 1.9 eV (Fig. \ref{fig8}) , as a result of the cross sections for the $^3A'$ and $^3A''$ PES having a maximum at 2.1 and 1.7 eV, respectively. This is in contrast with the $J$-shifting calculation which continues to increase and possibly peaks at higher collision energies. As explained in the previous section, we believe that our results describe more accurately the reaction at higher energies since the $J$-shifting approximation is too simplistic to correctly capture the relevant physics. 

% \subsection{Rotational and vibrational excitation of OH}

Rotationally resolved cross sections for selected collision energies of 0.781 eV (18 kcal/mol) and 2.5 eV are given in Fig. \ref{fig9}. The first energy can be compared directly to the existing results\cite{2004JChPh.121.6346B,2004JChPh.120.4316B}, and the second is of particular interest in non-thermal dynamics of hot O atoms and OH molecules in martian atmosphere\cite{2005SoSyR..39...22K}.
Our rotationally resolved cross sections for the production of OH($v'=0-1,j'$) molecules by vibrationally excited H$_2(v=1,j=0)$ (Fig. \ref{fig9}$(b)$) agree well with QCT results of Braunstein \textit{et al.}\cite{2004JChPh.120.4316B}, for which the highest excited rotational level, $j'_{\mathrm{max}}=25$ for OH$(v'=0)$ and $j'_{\mathrm{max}}=20$ for OH$(v'=1,j')$, fully agree, while the cross sections' values agree within 10 \%. Therefore, we feel that our results support the analysis of the OH$(v',j')$ state distributions based on the information theory\cite{2004JChPh.120.4316B}.
Balakrishnan\cite{2004JChPh.121.6346B} predicted rotationally resolved cross sections to be more than an order of magnitude smaller, with $j'_{\mathrm{max}}=15$ for OH$(v'=0)$ and $j'_{\mathrm{max}}=10$ for OH$(v'=1,j')$. Since a similar QM approach was used in both studies, we believe that these differences are due to the choice of basis set and larger rotational quantum number cutoff value used in this work.

\begin{figure*}[htb]
\centering
   \begin{tabular}{@{}cc@{}}
      \includegraphics[width=.5\linewidth]{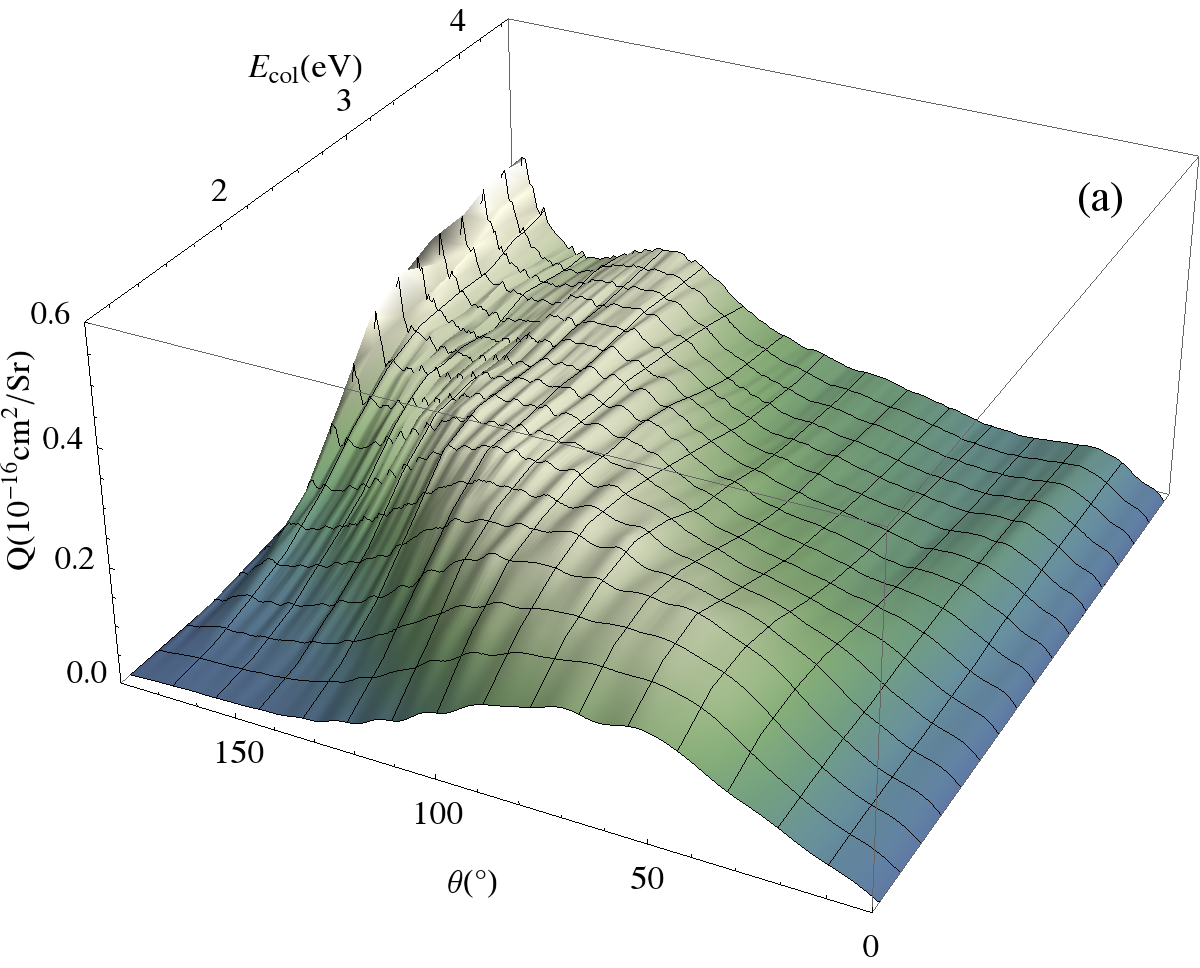} &
      \includegraphics[width=.5\linewidth]{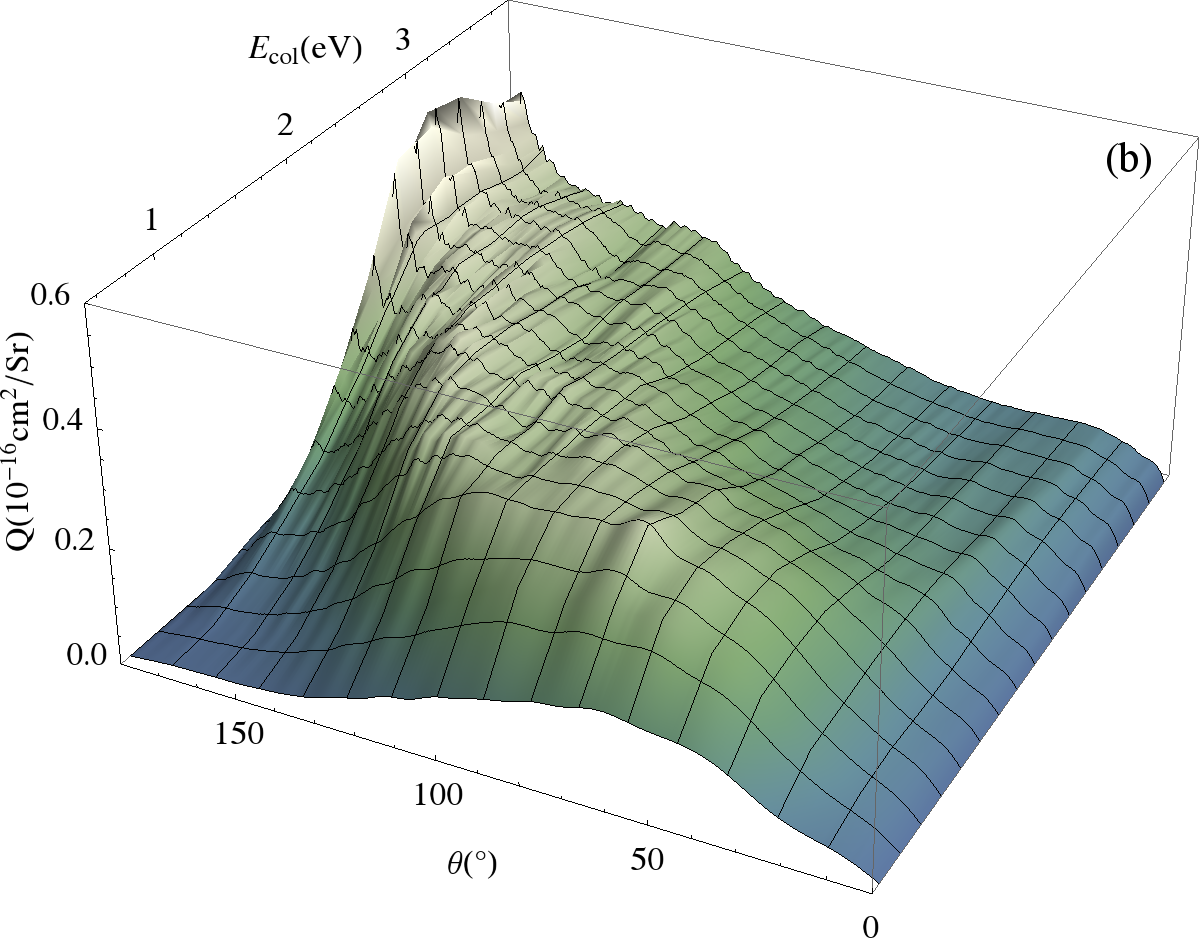} \\
      \includegraphics[width=.5\linewidth]{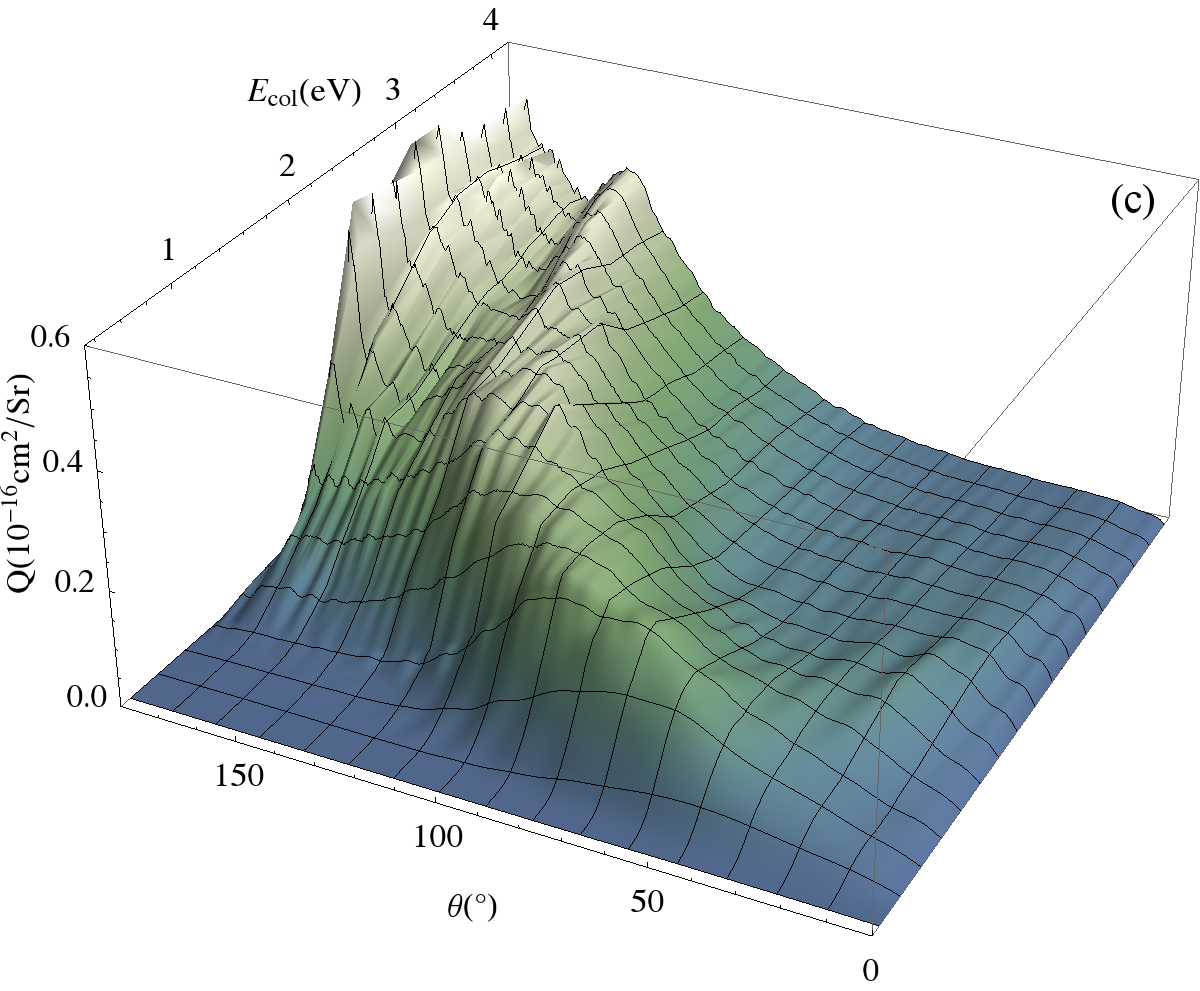} &
      \includegraphics[width=.5\linewidth]{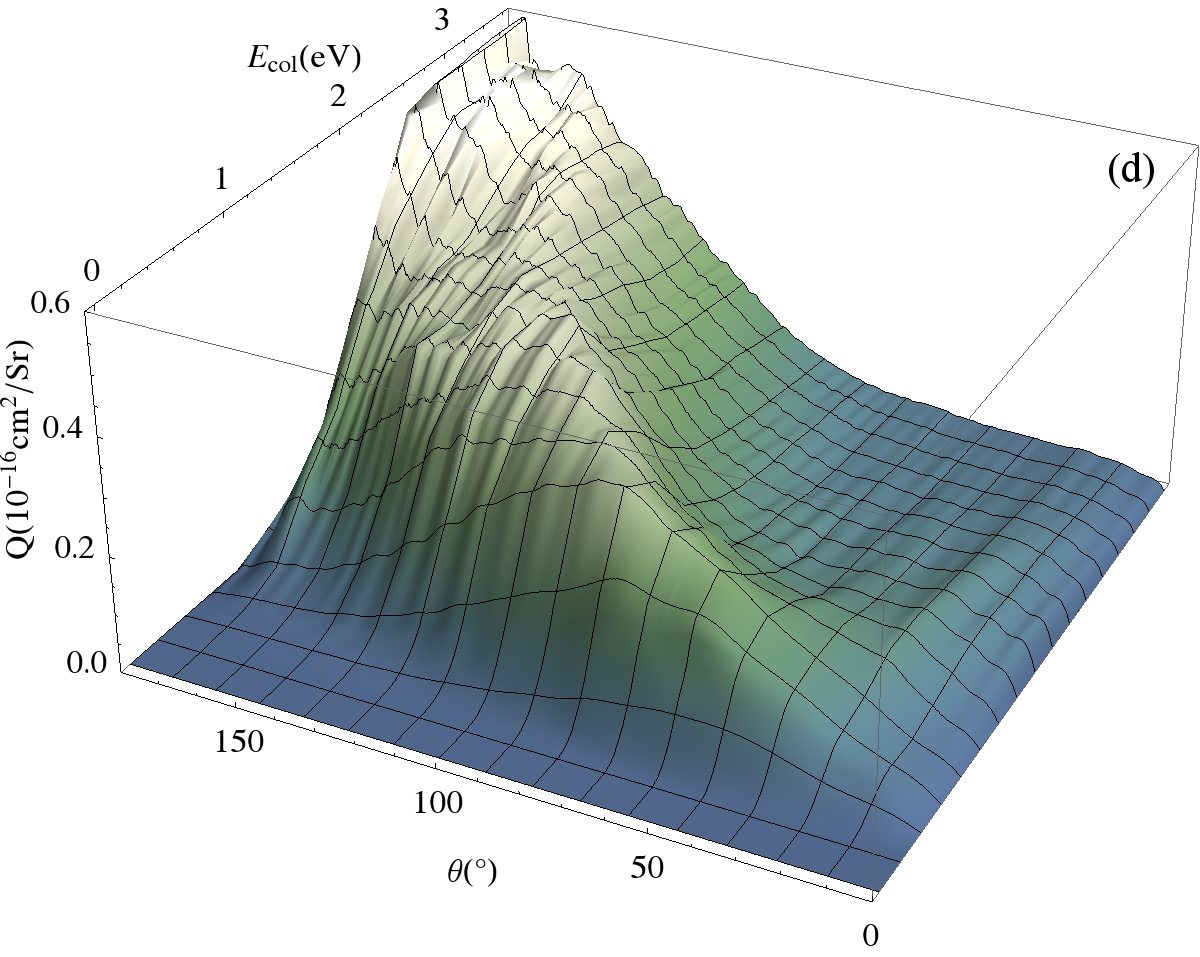} \\
   \end{tabular}
   \caption{\label{fig11} Total (summed over all product states OH$(v',j')$) differential cross sections $Q(\theta,E) = \sin \theta \, \mathrm{d} \sigma_{vj}(\theta,E) / \mathrm{d} \Omega$ as functions of the scattering angle $\theta$ and collision energy for O$(^3P)$+H$_2(v,j=0) \rightarrow$ OH$(v',j')$ + H reaction. The DCSs are shown separately for $^3A'$ (top panels) and $^3A''$ (bottom panels) PESs, and initial vibrational states $v=0$ (a,c) and $v=1$ (b,d).}
\end{figure*}

Fig. \ref{fig10} shows rotationally resolved cross sections for production of OH$(v'=0-1,j')$ molecule in O$(^3P)$+H$_2(v=0-1,j=0)$ reactive collision for collision energies from 0.6 to 4.1 eV. The rotational distributions peak at higher rotational quantum number as the collision energy increases (Fig. \ref{fig10}$(a,b)$). For the initial vibrational quantum number $v=1$ (Fig. \ref{fig10}$(c,d)$) more internal energy is available for exciting rotational degrees of freedom of the product OH molecule, resulting in broader rotational distributions shifted towards higher rotational states. In particular, in case of vibrational relaxation from $v=1$ to $v'=0$ (Fig. \ref{fig10}$(c)$), we obtain a double-peak distribution with respect to rotational states $j'$ of the product molecule as the internal energy release during the reaction becomes sufficient to excite higher rotational quantum numbers.
A similar behavior of rotational excitations in the reaction has been predicted in previous studies\cite{2004JChPh.121.6346B,2004JChPh.120.4316B}.

\subsection{Differential cross sections}

Fig. \ref{fig11} shows differential cross sections (DCSs) summed over all product states for the O$(^3P)$+H$_2(v,j=0) \rightarrow$ OH + H reaction, where $v=0,1$, as functions of scattering angle $\theta$ and collision energy. The scattering angle $\theta$ is defined as the angle between the velocity vector of the produced OH molecule and the velocity vector of the incoming O atom, where the vectors are given in center-of-mass frame. Therefore, $\theta=0$\textdegree \, corresponds to forward scattering and $\theta=180$\textdegree \, to backscattering. The DCSs are shown separately for the $^3A'$ and $^3A''$ surface and multiplied by the factor $a = \sin (\theta)$ to simplify the presentation for small forward and backward scattering angles and to highlight the region where the momentum transfer to the product OH molecule is the largest.  

For $^3A''$ PES and $v=0$ (Fig. \ref{fig11}$(c)$) we can identify main features shared between all shown cases. At low collision energies, below about $0.8$ eV, the DCS starts to form a small extended maximum between $\theta=10$\textdegree \, and $\theta=90$\textdegree, peaking at about $\theta=50$\textdegree$-70$\textdegree. As the collision energy increases, the maximum moves smoothly to larger scattering angles and becomes more pronounced, forming a central crescent-shaped ``ridge structure'' centered at about $\theta=120$\textdegree \, and $1.8 < E < 2$ eV, and peaking for $E > 3.8$ eV at about $140$\textdegree$<\theta<150$\textdegree. 
At higher collision energies, $E > 2$ eV, the backscattering contribution starts to dominate and forms a double-maximum structure with the ``ridge''. 
The presence of the ``ridge'' structure indicates a strong preference for sideways scattering of the product OH molecule. Similar structure of the reactive DCS was observed by Han \textit{et al.}\cite{JCC:JCC21940} for the collision energy of 1.2 eV.
Resonance oscillations due to interferences are visible in DCSs as linear ``ripples'' for selected scattering angles. For small angles, starting from about $\theta=20$\textdegree \,, they appear as simple evenly-spaced lines, while for larger scattering angles they become denser and better defined.
Analogous features, including the ridge structure and resonance oscillations, were found in H+D$_2$ reactive scattering\cite{2001JChPh.114.8796K}, where the author analyzed DCSs for a number of $(v',j')$ states of the product HD molecule. 

By performing a similar analysis of state-to-state scattering DCSs\cite{misc_detailedDCS}, we found that the width of the ridge, for each energy and the vibrational product state $v'$, is determined by a distribution of the DCSs for rotational product states $j'$, where higher rotational states extend the ridge towards lower scattering angles. 
Higher vibrational product states $v'$ tend to produce comparable but flatter distributions with a stronger backscattering component, shifted to smaller scattering angles by $10-40$\textdegree \, per $v'$, depending on the collision energy. The ridge itself is formed mostly by the contributions from $v'<3$ product states. These findings are in good agreement with the reactive production of HD \cite{2001JChPh.114.8796K}.

For the reactant H$_2$ initially in $v=1$ state, the DCSs remind of the $v=0$ case, although the double-maximum structure is less pronounced and present for a smaller range of collision energies, $1.6<E<2.4$ eV. 
Vibrational energy of the reactant molecule transfers more efficiently to higher rotational product states $j'$, forming a broader but shallower distribution (see Figs. \ref{fig9} \& \ref{fig10}). This reflects in DCSs as an increase in the width of the ``ridge'' for lower angles, and lowered amplitudes for higher angles. A comparable result was obtained using QCT for the same system\cite{2004JChPh.120.4316B}.

The described features are also present for $^3A'$ PES (Fig. \ref{fig11}, \textit{top panels}). The central ridge retains its crescent shape, but it is flatter and more extended with respect to the scattering angle than for $^3A''$ PES. This is particularly visible for collision energies $E>2$ eV and $\theta \approx 150$\textdegree \,, where the ridge remains largely flat for $^3A'$ PES as the energy increases, in contrast to the sharp peak present for $^3A''$ PES. The resonance oscillations remain largely unchanged and a statistically-weighted average taken over both surfaces does not cause them to disappear.

\subsection{Momentum transfer cross sections}

\begin{figure}[t]
\includegraphics[width=20pc]{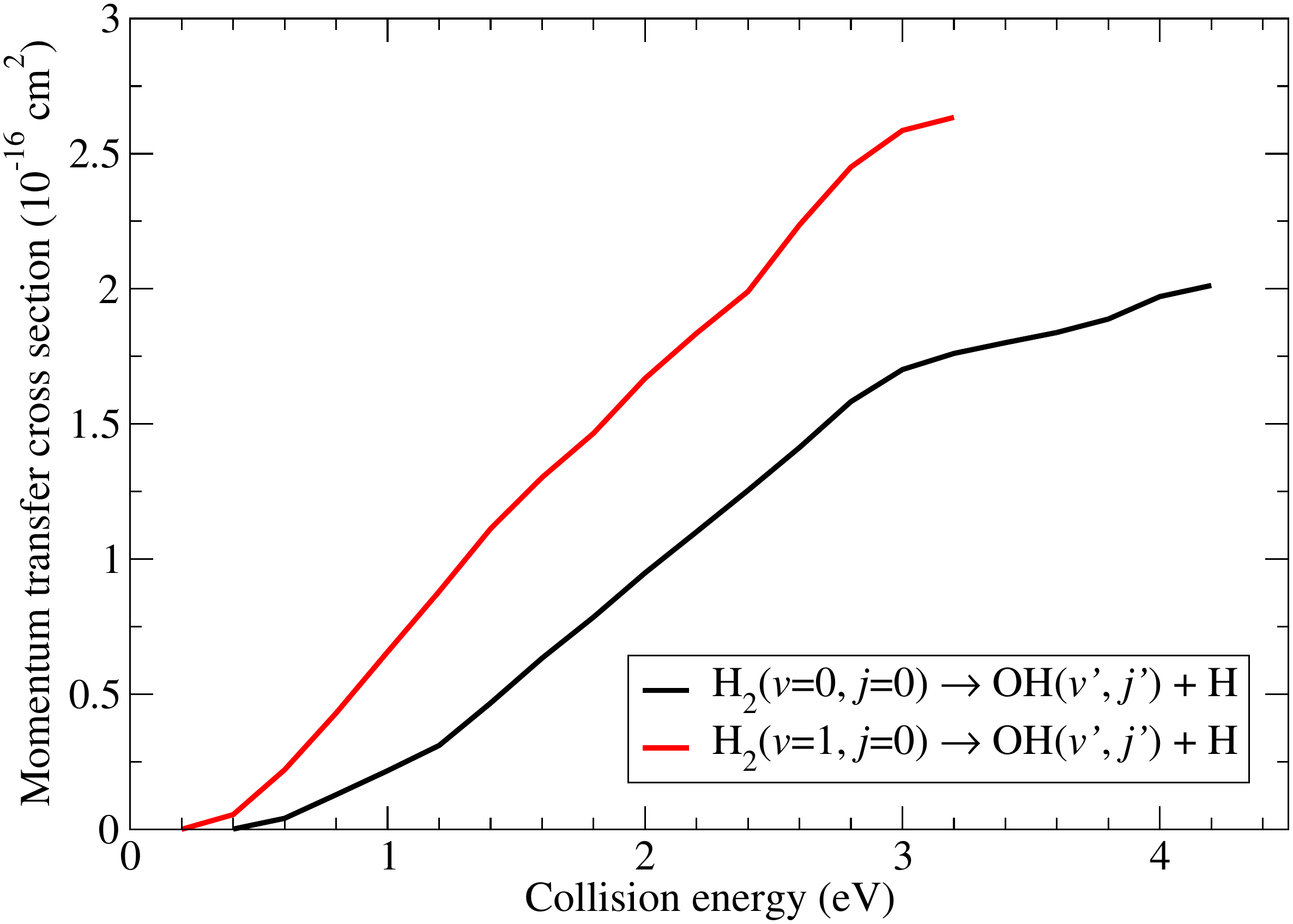}
\caption{\label{fig12} Total momentum transfer cross sections the reaction O$(^3P)$+H$_2(v=0-1,j=0)$.}
\end{figure}

Momentum transfer (diffusion) cross section is an effective cross section used to describe the average momentum transferred from a projectile to the target particle, often used in energy exchange and transport studies in astrophysics and atmospheric physics\cite{1962PApPh..13..643D}. 
Momentum transfer cross sections for the momentum transferred to the OH molecule in the reaction O$(^3P)$+H$_2(v,j=0) \rightarrow $ OH$(j',v')$ + H were evaluated using Eq. (\ref{eq3}) and statistically averaged for the two PESs in the same way as the total cross sections. Their values for the sum over all product states OH$(v',j')$, for two lowest vibrational states of the H$_2$ molecule, $v=0,1$, are given in Fig. \ref{fig12} and Table \ref{table2}\cite{misc_detailedDCS}.
Note that the ratio of the momentum transfer cross sections to the integral cross section for the reaction changes from greater than 4 to about 1 with increasing energy, unlike for the elastic process where the ratio is largely constant\cite{2011GeoRL..3802203B}. This reflects the preference for the OH molecule to be produced for ``sideways'' scattering angles (Fig. \ref{fig11}). It also indicates that the momentum transfer up to five times as large as for the elastic process can occur in the reaction at the high end of the considered collision energy range.

\begin{table}[tb]
\caption{\label{table2}Total momentum transfer cross sections for O$(^3P)$+H$_2(v,j=0)$ reaction, where $v=0,1$.}
\begin{ruledtabular}
\begin{tabular}{ccc}
   $E$(eV) & $\sigma^{\mathrm{mt}}_{v=0,j=0}$ $(10^{-16}$ cm$^2$) & $\sigma^{\mathrm{mt}}_{v=1,j=0}$ $(10^{-16}$ cm$^2$) \\
   \hline
    0.2 &      & 0.0002 \\
    0.4 & 0.001 & 0.05 \\
    0.6 & 0.04 & 0.22 \\
    0.8 & 0.12 & 0.43 \\
    1.0 & 0.21 & 0.66 \\
    1.2 & 0.31 & 0.88 \\
    1.4 & 0.47 & 1.11 \\
    1.6 & 0.63 & 1.30 \\
    1.8 & 0.79 & 1.47 \\
    2.0 & 0.95 & 1.67 \\
    2.2 & 1.10 & 1.84 \\
    2.4 & 1.25 & 1.99 \\
    2.6 & 1.41 & 2.24 \\
    2.8 & 1.58 & 2.45 \\
    3.0 & 1.70 & 2.59 \\
    3.2 & 1.76 & 2.36 \\
    3.4 & 1.80 & 2.59 \\
    3.6 & 1.84 & \\
    3.8 & 1.89 & \\
    4.0 & 1.97 & \\
    4.2 & 2.01 & \\
    4.4 & 1.95 & \\
\end{tabular}
\end{ruledtabular}
\end{table}

\section{Summary and Conclusions}

We have conducted a state-to-state quantum mechanical study of O$(^3P)$+H$_2 \rightarrow$ OH + H reaction using recent, chemically accurate $^3A'$ and $^3A''$ potential energy surfaces of Rogers \textit{et al.}\cite{Rogers_Kupperman_PES_2000} and Branda\~o \textit{et al.}\cite{2004JChPh.121.8861B_PES}, respectively. The calculation was carried out for collision energies of interest in astrophysics and atmospheric chemistry, ranging from 0.4 to 4.4 eV in the center-of-mass frame, with focus on the two lowest vibrational states of the H$_2$ molecule. 
The total reactive cross sections were constructed using thirty energy points per surface for each value of the total angular momentum up to $J_\mathrm{max}=105$.

The cross sections obtained in this study for the reactive production of OH molecules generally agree well with previous studies at low collision energies, up to about $E<0.7$ eV. At higher energies, where $0.7<E<1.3$ eV, our cross sections are larger by about 10-15 \% than those obtained in similar studies, with the exception of OH($v'=1$) where the agreement remains excellent. Another notable exception is the QCT study of Han \textit{et al.} that agrees well with our calculation for $E>0.9$ eV. In case of the $^3A''$ surface, our results appear to match recent QCT calculations of Yu-Fang \textit{et al.}\cite{1674-1056-20-7-078201} and Wang \textit{et al.}\cite{2006CPL...418..250W}, up to a small positive offset. 
For $E>1.6$ eV, regardless of the vibrational excitation of the reactants, we have obtained significantly smaller and qualitatively different reactive cross sections than predicted by the $J$-shifting approximation\cite{2005ApJ...629..305S}. This result, in particular, may be important for reaction rates with vibrationally-excited molecular hydrogen in astrophysical photon-dominated regions\cite{2010ApJ...713..662A}.

For the O$(^3P)$+H$_2(v=1)$ collisions, our cross sections are larger than predicted by the QM calculation of Sultanov \textit{et al.}\cite{2005ApJ...629..305S}, and appear to be in good agreement with the QCT calculation of Braunstein \textit{et al.}\cite{2004JChPh.120.4316B}.
We believe that the differences between our results and existing QM studies are mostly caused by differences in size and truncation of the bases used in the calculations. To test this, we have carried out a calculation on a reduced basis set using the truncation parameters of Balakrishnan\cite{2004JChPh.121.6346B} and obtained an almost complete match of the partial and total cross sections for the $^3A'$ surface of Rogers et al.\cite{Rogers_Kupperman_PES_2000}. Another possible source of disagreement are the differences, specifically in the van der Waals region\cite{B608871F}, between the Rogers' $^3A''$ \cite{Rogers_Kupperman_PES_2000} and the more recent BMS1 surface\cite{2004JChPh.121.8861B_PES}.

In addition, we have calculated differential cross sections and momentum transfer cross sections for the reaction, where the reactant H$_2$ was assumed to be in one of the two lowest vibrational states, $v=0,1$. For intermediate energies considered in this study, the DCSs indicate a strong preference for the reaction to proceed sideways, \textit{i.e.} for the product OH molecule to scatter perpendicular to the direction of propagation of the products, while for the energies greater than about 2 eV the backscattering starts to dominate. 
We expect these results to be of particular interest in studies of energy transport and dynamics in planetary and exoplanetary atmospheres\cite{2008SSRv..139..355J}, interstellar gas chemistry\cite{2010ApJ...713..662A}, and related astrophysical environments.

% If in two-column mode, this environment will change to single-column format so that long equations can be displayed. 
%\begin{widetext}
%$$\mbox{put long equation here}$$
%\end{widetext}

% Figures should be put into the text as floats. 
% Use the graphics or graphicx packages (distributed with LaTeX2e).
% Put the label that you will use with \ref{} command in the braces of the \label{} command.
%
% \begin{figure}
% \end{figure}

% \begin{table}
% \end{tabular}
% \end{table}

\begin{acknowledgments}
We are grateful to the authors of the respective studies for providing the Fortran subroutines for calculating the potential energy surfaces. We thank P. Zhang and N. Balakrishnan for useful discussions. This work was supported by NASA grant NNX10AB88G.
\end{acknowledgments}

% \bibliographystyle{agu08}

\bibliography{OH2}

\begin{thebibliography}{67}
\bibitem[{mis()}]{misc_detailedDCS}
\bibinfo{howpublished}{Results of the state-to-state analysis are available on
  request.}
\end{thebibliography}

\end{document}